\documentclass[reprint, amsmath,amssymb,aps, prl,superscriptaddress]{revtex4-1}
\usepackage{amsfonts,amssymb}
\usepackage{graphicx}
\usepackage{hyperref}    
\usepackage{bm}          
\usepackage{soul} 
\usepackage{color}
\usepackage{natbib}
\usepackage{longtable}
\usepackage{ textcomp }

\begin{document}

\author{Ting Chen}
\affiliation{Department of Chemical Engineering and Materials Science, University of Minnesota,  Minneapolis, Minnesota 55455, United States} 
\author{K. V. Reich}
\email{kreich@umn.edu}
\affiliation{Fine Theoretical Physics Institute, University of Minnesota, Minneapolis, MN 55455, United States}
\affiliation{Ioffe Institute, St. Petersburg, 194021, Russia}
\author{Nicolaas J. Kramer}
\affiliation{Department of Mechanical Engineering, University of Minnesota, Minneapolis, Minnesota 55455, United States}
\author{Han Fu}
\affiliation{Fine Theoretical Physics Institute, University of Minnesota, Minneapolis, MN 55455, United States}
\author{Uwe R. Kortshagen}
\affiliation{Department of Mechanical Engineering, University of Minnesota, Minneapolis, Minnesota 55455, United States} 
\author{B. I. Shklovskii}
\affiliation{Fine Theoretical Physics Institute, University of Minnesota, Minneapolis, MN 55455, United States}

\title{Metal-insulator transition in films of doped semiconductor nanocrystals}

\begin{abstract}
To fully deploy the potential of semiconductor nanocrystal films  as low-cost electronic materials, a better understanding of the amount of  dopants required to make their conductivity metallic is needed. In bulk semiconductors, the critical concentration of  electrons at the metal-insulator transition is described by the Mott  criterion. Here, we theoretically derive the critical concentration $n_c$ for  films of heavily doped nanocrystals devoid of ligands at their surface and  in direct contact with each other. In the accompanying experiments, we  investigate the conduction mechanism in films of phosphorus-doped,  ligand-free silicon nanocrystals. At the largest electron concentration  achieved in our samples, which is half the predicted $n_c$, we find that the localization length of hopping electrons is close to three times the  nanocrystals diameter, indicating that the film approaches the  metal-insulator transition.
\end{abstract}

\maketitle

\bigskip
\bigskip
Semiconductor nanocrystals (NCs) have shown great potential in optoelectronics applications such as solar cells \cite{gur_air-stable_2005}, light emitting diodes \cite{wood_colloidal_2010}, and field-effect transistors \cite{turk_gate-induced_2014,gresback_controlled_2014} by virtue of their size-tunable optical and electrical properties \cite{alivisatos_semiconductor_1996} and low-cost solution-based processing techniques \cite{sargent_infrared_2009,murray_synthesis_1993}. These applications require conducting NC films and the introduction of extra carriers through doping can enhance the electrical conduction. Several strategies for NC doping have been developed. Remote doping, the use of suitable ligands as donors in the vicinity of NC surface, increased the conductivity of PbSe NC films by 12 orders of magnitude\cite{talapin_pbse_2005}. Electrochemical doping, which tunes the carrier concentration accurately and reversibly, resulted in conducting NC films \cite{yu_n-type_2003,wang_electrochromic_2001}. Lately, stoichiometric control has emerged as a strategy to dope lead chalcogenide NCs \cite{oh_stoichiometric_2013}. Finally, electronic impurity doping of NCs, originally impeded by synthetic challenges \cite{norris_doped_2008}, was recently achieved in InAs \cite{mocatta_heavily_2011} and CdSe \cite{sahu_electronic_2012} NCs. 

While many experimental studies have been directed towards increasing the  conductivity of NC films, there is still no clear consensus on the fundamental question: what is the condition for the metal-insulator transition (MIT) in NC films \cite{guyot-sionnest_electrical_2012, shabaev_dark_2013, Band_like_transport_review}? In a bulk semiconductor, the critical electron concentration $n_M$ for the MIT depends on the Bohr radius $a_B$ according to the well-known Mott criterion \cite{mott_metal-insulator_1968} 

\begin{equation*}
n_M a_{B}^{3} \simeq 0.02,
\label{mott_eq}
\end{equation*} 

\noindent where $a_B=  \varepsilon\hbar^2/m^* e^2$ is the effective Bohr radius (in Gaussian units), $\varepsilon$ is the dielectric constant of the semiconductor, and $m^*$ is the effective electron mass. It is obvious that a dense film of undoped semiconductor NCs  is an insulator, while a film of touching metallic NCs with the same geometry is a conductor. Therefore, the MIT has to occur in semiconductor NC films at some critical concentration of electrons $n_c$, i.e. there should be an analog to the Mott criterion in a dense film of touching semiconductor NCs. 

\begin{figure}
\includegraphics[width=0.5 \textwidth ]{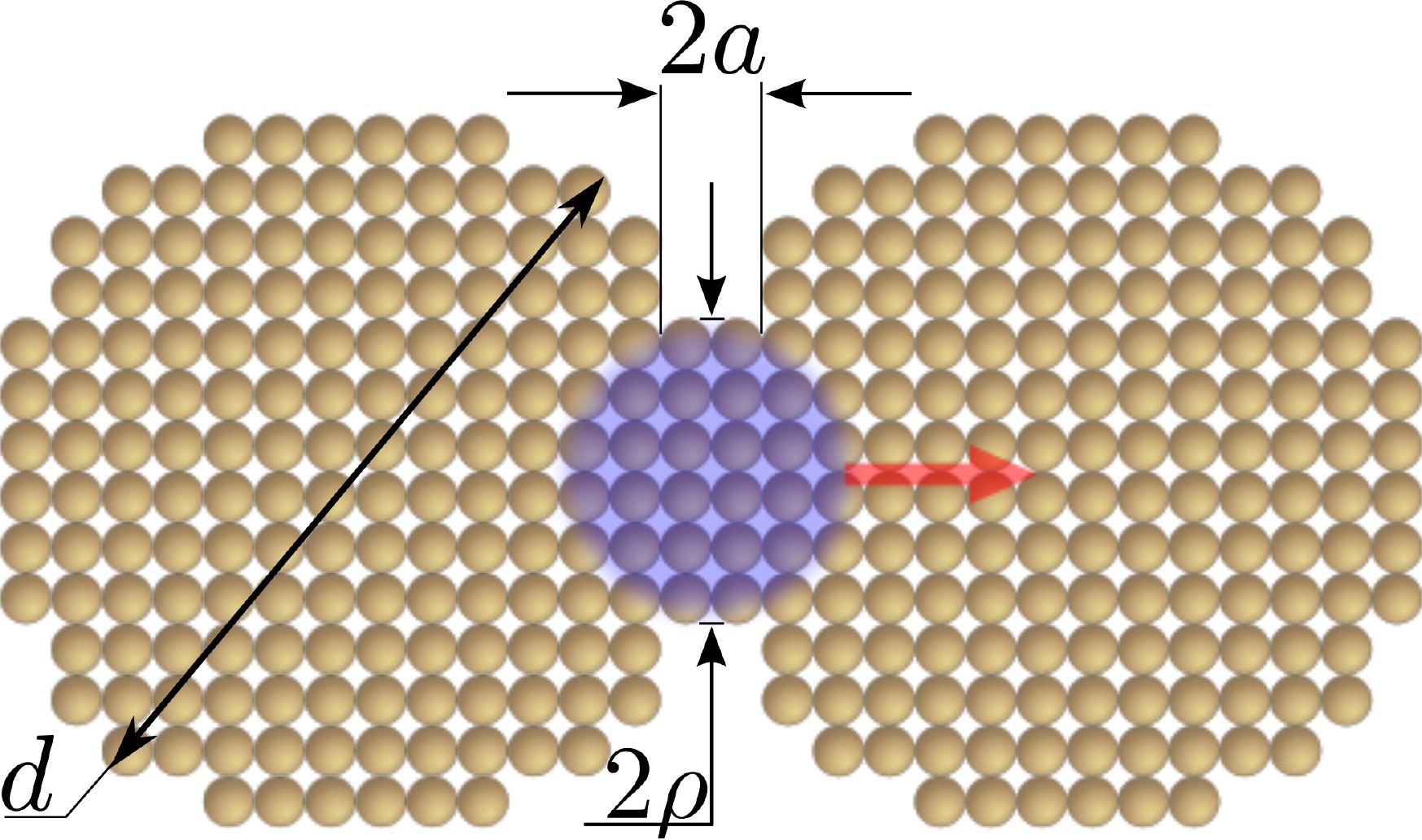}
\caption{{\bf The origin of the metal-to-insulator transition in semiconductor nanocrystal films.} The figure shows the cross section of two nanocrystals in contact through facets with radius $\rho$. The blue spherical cloud represents an electron wave packet which moves through the contact. Such a compact wave packet is available only at $k_F \rho > 2$ (see Eq. (\ref{eq:MIT_final}) and equivalent Eq. (\ref{eq:result})). Here $a$ is the lattice constant, $d$ is the NC diameter, $k_F$ is the Fermi wave vector.}
\label{fig:Sheme}
\end{figure}

Here, we focus on NCs that touch each other through small facets of radius $\rho$ without any ligands that impede conduction (Fig. \ref{fig:Sheme}). We derive below that for such touching NCs the MIT criterion is

\begin{equation}
  \label{eq:result}
  n_c \rho^3 \simeq 0.3 g,
\end{equation}
\noindent 
where $g$ is the number of equivalent minima in the conduction band of the semiconductor. As to be expected, Eq. (\ref{eq:result}) predicts $n_c$ for NC films that is much larger than $n_M$ for the bulk. For instance, for close to spherical particles, the facet radius imposed by the discretness of the crystal lattice can be approximated as $\rho_a=\sqrt{da/2}$, where $a$ is the lattice constant and $d$ is the NC diameter. For such facets and CdSe NCs with $d=5$ nm, Eq. (\ref{eq:result}) gives $n_c = 2 \times 10^{20} \mathrm{cm}^{-3}$, while Mott's criterion equation~(\ref{mott_eq}) yields $n_M = 2 \times 10^{17} \mathrm{cm}^{-3}$ for bulk CdSe. For an array of  Si NCs with  $d = 8~\mathrm{nm}$  we find $n_c \simeq 5 \times 10^{20} ~\mathrm{cm^{-3}}$, compared to $n_M=3\times 10^{18}~\mathrm{cm^{-3}}$ from the Mott criterion.

Below, we derive Eq.~(\ref{eq:result}) and discuss its applicability and limitations. To test the predictions of our theory, we investigate the electron transport in dense films of phosphorus-doped, ligand-free Si NCs over a wide range of doping concentration. We find that the electron localization length grows with $n$ and becomes 3 times larger than $d$ at $n \simeq 0.5 n_c$, where $n_c$ is predicted by Eq. (\ref{eq:result}). This signals that the MIT is indeed occurring close to predicted $n_c$.

\section{Critical doping concentration at MIT}

For metallic transport to occur in arrays of spherical NCs that touch each other at transport limiting facets, the NCs themselves need to be metallic, i.e. the number of electrons $N$ in a NC is large. Hence, the electron gas can be described with the Fermi wave vector:

\begin{equation}
  \label{eq:Fermi_wave}
k_{F}=\left(\frac{3 \pi^2}{g} n \right)^{1/3}.
\end{equation}
\noindent Here $n=6N/\pi d^3$ is the density of electrons in a NC. Below, $k_F$ serves as a measure of the concentration $n$.   In Ref. \cite{localization_length_NC} we show that if $d \gg a_B$ the NC has random energy spectrum filled upto $\epsilon_F=\hbar^2 k_F^2/2m^*$ due to random potential of donors. In the opposite quantum-confined case, $d \ll a_B$, for spherically symmetric NCs, electrons occupy states with different radial and angular momentum quantum numbers $(n,l)$-shells, each of them being degenerate with respect of azimuthal quantum number $m$. If the total number of electrons in the NC is $N_D  \gg 1$, several $(n,l)$-shells are occupied. Still, when quantum numbers are large, Bohr's correspondence principle allows us to consider the average density of states of electrons quasiclassically and introduce the Fermi wave vector $k_F$ and the Fermi energy $\epsilon_F$. This description is correct if the Fermi energy $\epsilon_F$ is a good estimate for the energy of the top shell. At the critical concentration $n_c =2\times 10^{20}~\mathrm{cm^{-3}}$ for CdSe NCs with diameter $d=5~\mathrm{nm}$, each NC has 13 electrons and the top shell is the half-filled 1d-shell. The  Fermi energy $\epsilon_F$ at the concentration $n_c$ is 50\% smaller than the shell energy $\sim 60 \hbar^2/md^2$. Hence, our degenerate gas description is accurate within 50\%, which is a measure for the accuracy of our $n_c$ predictions.

To derive the MIT condition, we consider the conductance of two metallic NC connected by a small facet contact. When $k_F \rho \gg 1$, the conductance of such a ``point'' contact was previously studied quasiclassically~\cite{Sharvin,transport_through_constriction}:

\begin{equation}
  \label{eq:Sharvin_2}
  G= \frac{e^2}{4\pi^2 \hbar} k_F^2 \pi \rho^2,
\end{equation}

\noindent where $\pi \rho^2$ is the contact area. This conductance can be easily understood with the help of  the Landauer formula \cite{nazarov2009quantum}. The number of conducting channels in the contact area is $\sim (k_F\rho)^2$ and each of them additively contributes $\sim e^2/\pi \hbar$ to $G$. 

It was proven that the MIT occurs if the average conductance between two neighboring NCs $G$ in an array of NCs is equal to the minimal conductance $G_m$ \cite{Matveev,beloborodov_granular_2007}:

\begin{equation}
   \label{eq:minimal_G}
 G= G_m \equiv \frac{e^2}{\pi \hbar}.
\end{equation}

Substituting $G$ from Eq. (\ref{eq:Sharvin_2}) into  (\ref{eq:minimal_G}) yields the general criterion for the MIT

 \begin{equation}
   \label{eq:MIT_final}
  k_F \rho \simeq 2,
 \end{equation}
which can easily be rewritten in terms of $n_c$ with help of Eq. (\ref{eq:Fermi_wave}) to yield Eq.~(\ref{eq:result}) . The origin of Eq. (\ref{eq:MIT_final}) is illustrated in Fig. \ref{fig:Sheme}: the condition $k_F\rho>2$ describes electron wave packets with a size small enough to pass through the contact facet. 

We now discuss the effect of the contact facet size, which for metal chalcogenide NCs can be large \cite{facets}. Considering as an example an octahedron-shaped particle that is circumscribed by a sphere of diameter $d$, the area of each facet is $0.2d^2$. Equating this to $\pi \rho_1^2$, we find an effective $\rho_1 \simeq 0.26 d.$ At $d=8~\mathrm{nm}$, we get  $\rho_1 = 20~ \mathrm{\AA}$, which is not far from  $\rho_a=\sqrt{da/2}=14~\mathrm{\AA}$ for the spherical case. This $\rho_1$ results in a 3 times smaller $n_c$ than $\rho_a$. 

The NC contact through facets is, of course, the best case scenario that defines the lower bound for $n_c$ for spherical NCs. For NCs that do not touch through facets, a finite tunneling distance $b=\hbar/\sqrt{2mU_0}$ in the medium between NCs should be taken into account. Here $U_0$ is the work function. An electron can move between neighboring NCs only in a disc that we call $b$-contact, in which the distance between NCs is smaller than $b$. The radius of such a $b$-contact is $\rho_b=\sqrt{db/2}$ (See SI1 \ref{sec:b_contact}). The small ratio of the effective electron mass in the semiconductor NC to the free electron mass makes the $b$-contact transparent (see similar effect in Ref. \onlinecite{shabaev_dark_2013}).  Usually, for NCs in vacuum (air), $b \simeq 1 ~\mathrm{\AA}$ which is much smaller than the lattice constant $a$; relying on only $b$-contacts increases $n_c$ upto 10 times. 

So far we have considered NCs with bare surfaces. If NCs are covered by a thin shell of ligands or oxide leading to a NC separation $s$, the conductance $G$ acquires an additional factor  $\exp(-2s/b)$ and Eqs. (\ref{eq:minimal_G}),~(\ref{eq:MIT_final}) yield

\begin{equation}
   \label{eq:n_exp}
   n_c(s) \simeq n_c \exp\left( \frac{3 s}{b} \right).
 \end{equation}
In this case, the MIT may become unreachable. 

We also can calculate the low temperature mobility  $\mu$ in the vicinity of the MIT. Substituting the conductivity from Eq. (\ref{eq:Sharvin_2}) into the expression   $\mu=6G/\pi e n d$ (where the factor $6/\pi$ accounts for the difference between the concentration of electrons inside NCs and the average concentration in the film)  we find  the low-temperature metallic mobility
\begin{equation*}
  \label{eq:mobility_general}
\mu= \frac{3^{5/3}}{2 \pi^{2/3}} \frac{e}{\hbar}  \frac{\rho^2}{g^{2/3} n^{1/3} d}.
\end{equation*}
For CdSe NCs with $d=4~\mathrm{nm}$, $\mu$ is on the order of $10 ~\mathrm{cm^2/V \cdot s}$ at $n=2n_c$ and is close to the experimentally observed room-temperature mobility of $30 ~\mathrm{cm^2/V \cdot s}$ for CdSe ~\cite{lee_band-like_2011,choi_bandlike_2012, Talapin_2013_high_mobility}. Note that this mobility mostly is due to the contact resistance, while the  in  the case of bulk semiconductors the low temperature mobility is due to  the scattering by donors.

We now discuss the role of disorder. The number of donors $N_D$ in a NC randomly fluctuates between NCs with a Gaussian distribution. If each NC were neutral ($N=N_D$), Gaussian fluctuations of donor number, $\delta N_D$, would lead to substantial fluctuations $\epsilon_F/\sqrt{N_D}$ of $\epsilon_F$ from one NC to another. To establish a unique chemical potential of electrons (the Fermi level), electrons move from NCs with larger than average $n$ to ones with smaller than average $n$. Accordingly, most NCs attain net charges $\sim \sqrt{N}e$. This leads to large fluctuations of the Coulomb potential and moves the NC electron energy levels  with respect to the Fermi level. In an insulating NC array ($n < n_c$), this replaces the global charging energy gap of the density of states by  the Coulomb gap which leads to the Efros-Shklovskii variable range hopping~\cite{skinner_theory_2012} (see below).

This theory is based on the criterion (\ref{eq:minimal_G}), which guarantees that energy levels of NCs have a width comparable to the average energy difference between the adjacent levels $\delta$. This also eliminates the Anderson localization.  At the same time, at $G > G_m$, there is at least one  electron channel in the contact disk with almost perfect transparency. This guarantees~\cite{Matveev,nazarov2009quantum} that the charging energy of every single  NC, $E_c=e^2/\varepsilon_r d$, is reduced to a value much smaller than $\delta$ ($\varepsilon_{r}$ is effective dielectric constant of the NC film). Accordingly, the Mott-Hubbard localization is eliminated at the same time as the Anderson localization. We emphasize that criterion~(\ref{eq:minimal_G}) is universal and holds for heavily doped NC films regardless of whether NCs are quantum-confined or not. A detailed discussion of this universality is presented in SI2 \ref{sec:Anderson}.

Another generic disorder effect is the variation of NC sizes~\cite{shabaev_dark_2013}. Remarkably, in heavily doped NCs,  this variation does not lead to  Anderson localization of electrons, because their spectrum is already  random. Thus, as for contacting metallic NCs, small variations of the diameter are inconsequential. On the other hand, the $\approx$ 15\% size dispersion in our experiments below may complicate the matching of NC facets and therefore increase $n_c$.

\section{Electron Transport in Films of Heavily Doped Silicon Nanocrystals}

\bigskip
\bigskip

To test the predictions of our theory, we studied the electron transport in films of heavily phosphorous (P)-doped Si NCs.
Freestanding Si NCs were synthesized in a nonthermal radio-frequency  plasma reactor as reported previously \cite{mangolini_high-yield_2005}. We investigated six Si NC films with different  P concentrations. We refer to them using their nominal doping concentration $X_{P,nom}$, the fractional flow rate, defined as $X_{P,nom}= \mathrm{[PH_{3}]/([PH_{3}]+[SiH_{4}])}\times 100\%$, where [PH$_{3}$] and [SiH$_{4}$] are the flow rates of phosphine and silane.

\begin{table}[t!]
\label{tab1}
  \begin{tabular}{ c | c| c|  c|  c|  c }
    \hline
    $X_{P,nom} \%$ & $X_{P,ICP} \%$  & $d,~\mathrm{nm}$ & $F,~\mathrm{cm^{-1}}$ & $n,~10^{20}\mathrm{ cm^{-3}}$ & $\xi, ~\mathrm{nm}$ \\ \hline
    1 & 0.46 &8.1& - & - & 1.4 \\ 
    2 & 0.82 &8& - & - & 1.9 \\ 
    3 & 1.56 &8& - & - & 6.1 \\ 
    5 & 2.38 &8 &1110 & 1.9 & 12.7 \\ 
    10 & 4.06 &7.5 &1260 & 2.4 & 20.6 \\ 
    20 & 6.98 &7.1 &1360 & 2.8 & 26.8 \\ 
    \hline
  \end{tabular}
 \caption{{\bf Parameters of P-doped Si NCs.} $X_{P,nom}$ is the nominal doping, $X_{P,ICP}$ is the atomic fraction of P in Si NCs measured from ICP-OES, $d$ is the average diameter of NCs, $F$ is the plasmonic peak in wavenumber for Si NC films, $n$ is the electron concentration estimated from the plasmonic peak, $\xi$ is the localization length calculated from the electrical transport data.}
\end{table}

To quantify P incorporation in Si NCs, we utilized inductively coupled plasma optical emission spectroscopy (ICP-OES). Table I shows the P atomic fraction in Si NCs for each nominal doping concentration. We observe a monotonic increase in the incorporated P fraction with increasing nominal doping concentration, and the incorporation efficiency is about 50\% when $X_{P,nom}$ is below 5\%. However, this technique only measures the elemental composition but not the concentration of active dopants.

\begin{figure}[t!]
\centering
\includegraphics[width=0.4 \textwidth]{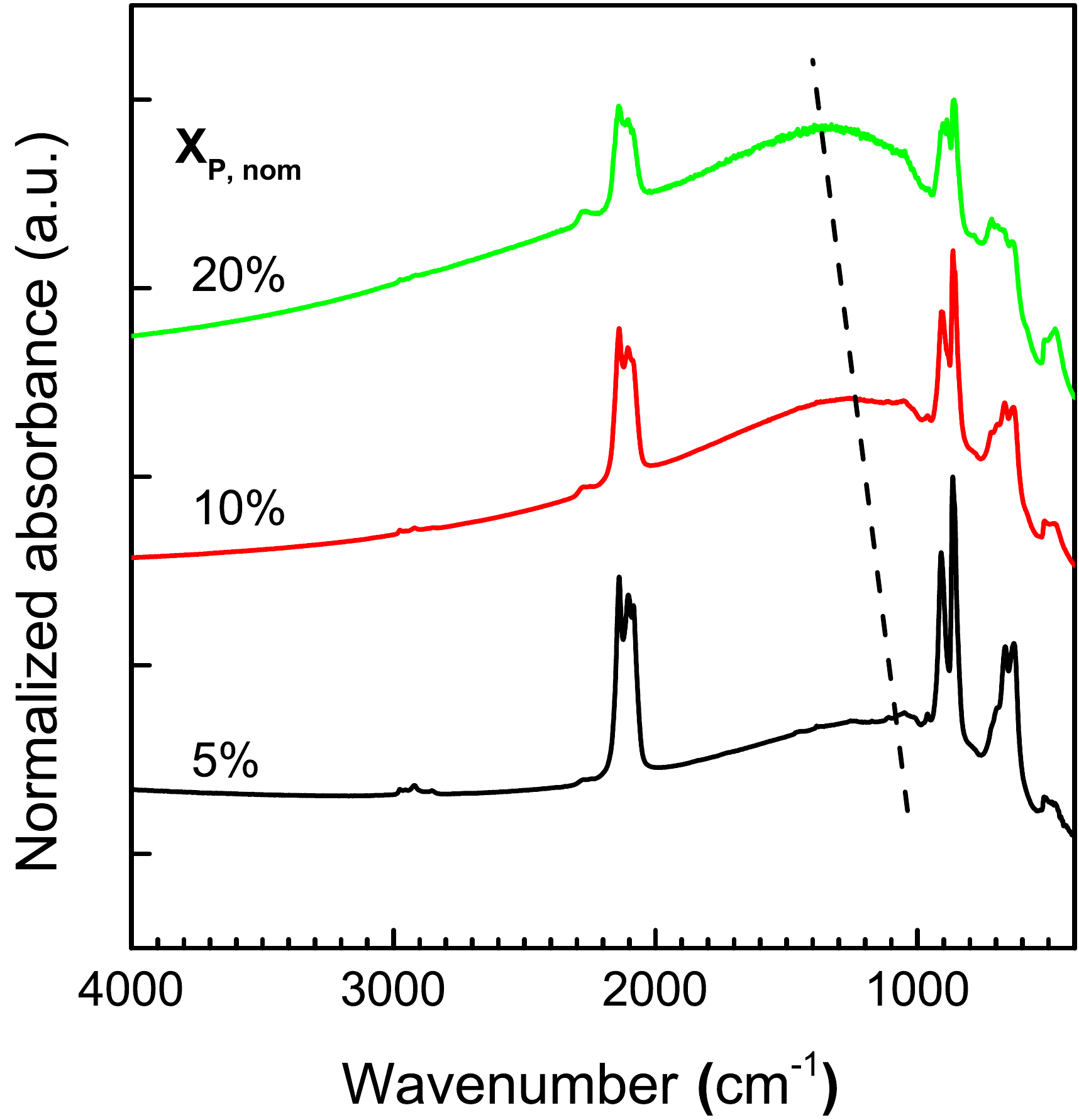} 
\caption{{\bf Determining the free carrier density from the localized surface plasmon resonance.} Fourier transform infrared (FTIR) spectroscopy spectra for nominal 5\%, 10\% and 20\% P-doped Si NCs. The broad absorption feature is the localized surface plasmon resonance. It shifts to higher wavenumbers with  increasing nominal doping concentration (a dashed line is added as a guide to the eye). The electron concentration is estimated from the plasmonic peak position and shown in the Table I. The sharp peaks around $2100~\mathrm{cm^{-1}}$ are associated with surface silicon hydride stretching modes, and features at $1000 - 800~\mathrm{ cm^{-1}}$ and $750 - 550~\mathrm{cm^{-1}}$ are the relevant deformation modes \cite{modes_phonon_Si}. The small peak at $2200 -2300~\mathrm{ cm^{-1}}$ can be either $\mathrm{Si-P_{x}-H_{y}}$ or $\mathrm{O-Si-H_{x}}$ since they appear in the same wavenumber range \cite{Si_H_modes}.}
\label{fig:concentration}
\end{figure}

For samples at sufficiently high doping concentration, we can determine the free electron concentration through the localized surface plasmon resonance (LSPR). The position of the plasmonic peak depends on the free electron concentration $n$ as described by the equation  \cite{rowe_phosphorus-doped_2013}

\begin{equation*}
\omega=\sqrt{\frac{4\pi n e^{2}}{m^{\ast }(\varepsilon+2\varepsilon _{m})}},
\label{eq1}
\end{equation*}

\noindent where $\omega$ is the localized surface plasmonic resonance frequency, $\varepsilon$ is the dielectric constant for bulk Si (11.7) and $\varepsilon_{m}$ is the dielectric constant for the surrounding medium, taken as $\sim$ 1 for nitrogen atmosphere in this study. As shown in Figure \ref{fig:concentration}, the plasmonic peaks are at 1110, 1260 and 1360 cm$^{-1}$ in the infrared absorption spectra and the free electron concentrations are $n=1.9\times 10^{20}, 2.4\times 10^{20}$, and $ 2.8\times 10^{20}$ cm$^{-3}$ for 5\%, 10\% and 20\% P-doped Si NCs, respectively. No plasmonic peaks were observed for doping concentrations lower than nominal 5\%. 
All known parameters of Si NCs are summarized in Table I. Si NC sizes were determined by X-ray diffraction and transmission electron microscopy, as presented in the SI3 \ref{sec:characterization}.  

\begin{figure}[t!]
\centering
\includegraphics[width=0.46 \textwidth]{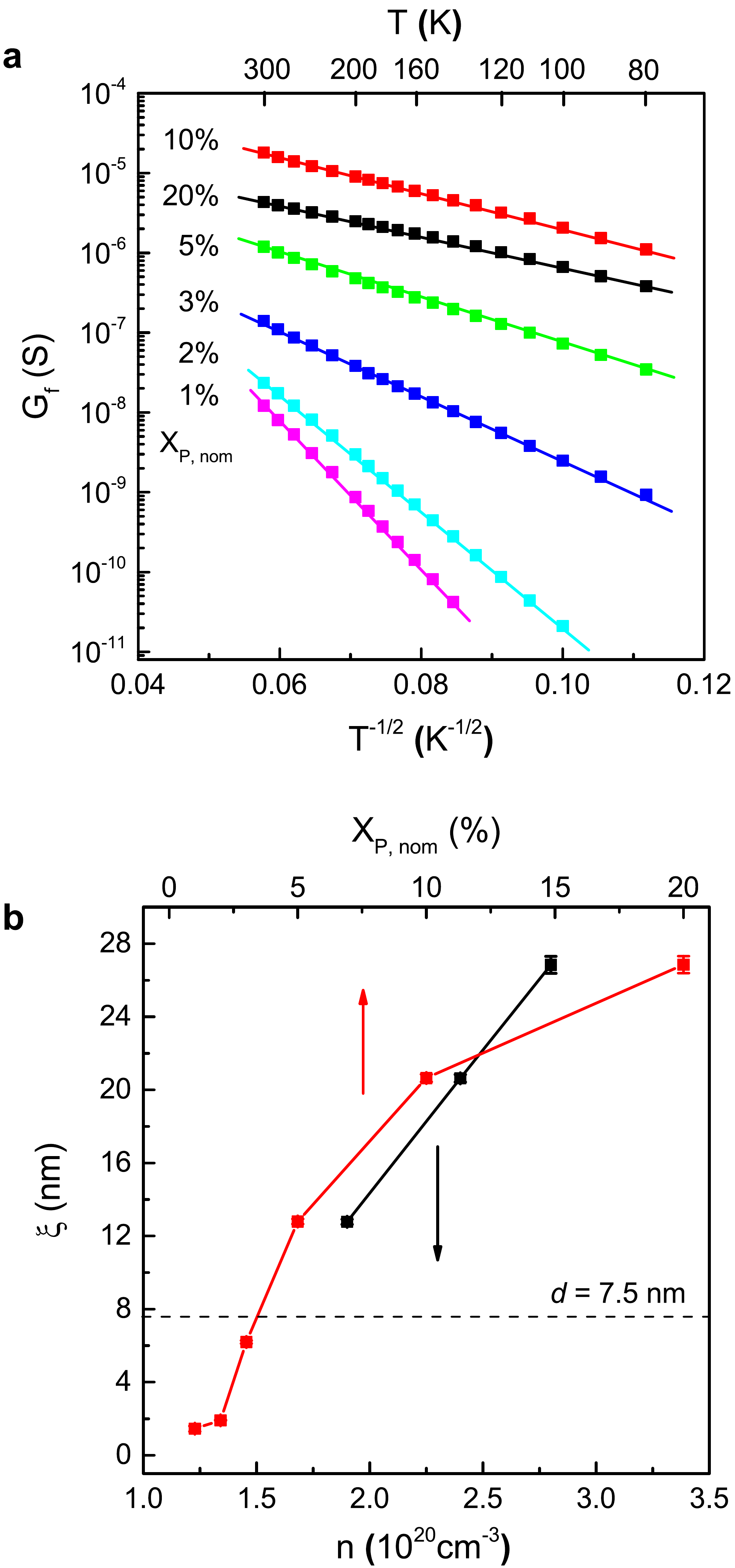} 
\caption{{\bf Electrical transport in phosphorous-doped Si NC films approaching the metal-to-insulator transition.} {\bf a} Temperature dependence of the ohmic conductance for films made from Si NCs at different nominal P doping concentrations. Solid lines are linear fits for each doping concentration. {\bf b} Localization length $\xi$ versus the electron concentration in a NC $n$ and the nominal  P doping concentration $X_{P,nom}$. Error bar for each $\xi$ comes from the uncertainty caused by linear fit and it is as large as the symbol size. The average diameter of a NC in films is shown by horizontal dashed line.}
\label{fig:G_T}
\end{figure}

Next, we examined the electrical transport in the P-doped Si NC films. Figure \ref{fig:G_T} depicts the temperature dependence of the ohmic conductance $G$ for P-doped Si NC films at nominal doping concentrations from 1\% to 20\%. As shown in Figure \ref{fig:G_T}a, the conductance of P-doped Si NC films monotonically increases with the nominal doping concentration. However, Si NC film at {\it X}$_{P,nom}$ = 20\% shows lower conductance than the film at {\it X}$_{P,nom}$ = 10\%. The reason is not clear at this time. 
Over the entire range of doping concentrations under investigation, the film conductance $G_f$ follows Efros-Shklovskii (ES) law:

\begin{equation}
  \label{eq:ES}
G_f  \propto  \exp \left[- \left(\frac{T_{ES}}{T}\right)^{1/2} \right],
\end{equation}
where 

\begin{equation}
T_{ES}=\frac{Ce^{2}}{\varepsilon _{r}k_{B} \xi}.
\label{eq3}
\end{equation}
\noindent Zabrodskii analysis \cite{zabrodskii_coulomb_2001} (not shown here) confirms this result.  The fact that ES conductivity is seen even at the smallest studied donor concentration implies that even in this case the average number $N_D$ of donors per NC is large. As  explained above, fluctuations of this number lead to charging of majority of NCs, the ES Coulomb gap and ES conductivity. 

We extract the characteristic temperature $T_{ES}$ from the slope of linear fits for $\ln G$ vs $ T^{-1/2}$ using Eq. \eqref{eq:ES}. We estimate the effective dielectric constant $\varepsilon_{r}$ of the NC film from the canonical Maxwell-Garnett formula \cite{maxwell_treatise_1881}(the film density is assumed to be $\sim$ 50\%) and find $\varepsilon_r \simeq 3$. Knowing $T_{ES}$, the dielectric constant and using Eq. \eqref{eq3} we compute the localization length  $\xi$. As the doping concentration increases, $\xi$ grows from 1.4 nm at $X_{P,nom}=1\%$ to 26.8 nm at $X_{P,nom}=20\%$, as displayed in Figure \ref{fig:G_T}b and shown in Table I.

For larger nominal doping $n =(1.9-2.8) \times 10^{20}~\mathrm{cm^{-3}}$, $\xi$ exceeds the NC diameter and reaches three NC diameters for the highest doping level. This indicates the approach to the MIT with growing $n$. These data are consistent with the predicted $n_c \simeq 5 \times 10^{20}~ \mathrm{cm^{-3}}$ by Eq.~(\ref{eq:result}). Similar growth of $\xi$ was observed in  Ref. \cite{turk_gate-induced_2014} for CdSe NCs.

We also studied the effect of the NC separation to verify  the theory prediction, Eq.~(\ref{eq:n_exp}). The P-doped Si NCs are prone to oxidation if exposed to air. An oxide shell starts to grow from the outer surface towards the core by consuming the original Si lattice. The neighboring NCs are now separated by two oxide shells, whose combined thickness $s$ grows with time. According to Eq. \eqref{eq:n_exp} the critical electron concentration $n_c$ increases with increasing $s$ and  therefore, $\xi$ decreases. In  SI4 \ref{sec:oxidation} we report our study of $\xi(s)$ and find a qualitative agreement with Eq. (\ref{eq:n_exp}).

\section{CONCLUSIONS}
We derived the MIT criterion given by Eq.~\eqref{eq:result} for films of semiconductor NCs analogous to the Mott criterion for bulk semiconductors.  According to this criterion, MIT occurs in Si NC films at a critical concentration  $n_c \simeq 5 \times 10^{20}~\mathrm{cm^{-3}}$.  We investigated the electron transport in P-doped Si NCs to test this theory. ES variable range hopping conduction was found  for all doping concentrations $n$ up to $n=2.8  \times  10^{20}~\mathrm{cm^{-3}}$. The localization length increases with increasing doping concentration and exceeds the diameter of a NC at $n>1.9 \times 10^{20}~\mathrm{cm^{-3}}$, which indicates the approach to the MIT in P-doped Si NC films, in agreement with our theory. 

Recently in the Ref. \cite{localization_length_NC} we focused on the variable-range hopping of electrons in semiconductor nanocrystal (NC) films below the critical doping concentration $n_c$ at which it becomes metallic.  We studied how the localization length grows with the doping concentration $n$ in the film of touching NCs. For that we calculated the electron transfer matrix element $t(n)$ between neighboring NCs. We used the ratio of $t(n)$ to the disorder-induced NC level dispersion to find the localization length of electrons due to the multistep elastic co-tunneling process and showed that the localization length diverges at concentration $n$  equal to $n_c$ given by Eq. \eqref{eq:result}.

\section{ACKNOWLEDGEMENT}

The authors would like to thank K.A. Matveev, C. Leighton, B. Skinner and A. Kamenev  for helpful discussions, C. D. Frisbie for the use of his equipment and R. Knurr for assistance with the ICP-OES analysis. Ting Chen (electrical transport studies)  and K.V. Reich(theory) were supported primarily by the National Science Foundation through the University of Minnesota MRSEC under Award Number DMR-1420013. Nicolaas Kramer(materials synthesis) was supported by the DOE Center for Advanced Solar Photophysics. Part of this work was carried out in the College of Science and Engineering Characterization Facility, University of Minnesota, which has received capital equipment funding from the NSF through the UMN MRSEC program. Part of this work also used the College of Science and Engineering Nanofabrication center, University of Minnesota, which receives partial support from NSF through the NNIN program.

\section{Author contribution}
K.V. Reich, Han Fu. and B.I. Shklovskii created the theory. Ting Chen performed the structural and electrical characterization, Nicolaas Kramer synthesized materials, Uwe R. Kortshagen discussed and supervised the work. All authors participated in the discussion and interpretation of the results and co-wrote the manuscript. 

\section{Competing financial interests}
The authors declare no financial interests.

\section{EXPERIMENTAL METHODS}
\label{sec:details} 
Freestanding P-doped Si NCs were synthesized in a nonthermal radio frequency plasma with a frequency of 13.56 MHz. The detailed description of synthesis can be found elsewhere \cite{mangolini_high-yield_2005,rowe_phosphorus-doped_2013}. The doping concentration is controlled by changing the flow rate of phosphine (PH$_{3}$) while maintaining constant flow rates for Ar and SiH$_{4}$ . Typical flow rates used in this work are 0.4 standard cubic centimeters per minute (sccm) of SiH$_{4}$, 55 sccm of Ar, and 0.028 - 0.66 sccm of PH$_{3}$ diluted to 15\% in hydrogen. The plasma is operated at a pressure of 0.9 Torr with a nominal power of 110 W.  

The crystallinity and the particle size of Si NCs were characterized by XRD using a Bruker-AXS microdiffractometer with a 2.2 kW sealed Cu X-ray source at 40 kV and 40 mA (wavelength 0.154 nm). The XRD pattern was recorded for dry powders of Si NCs deposited on a glass substrate. The high resolution bright field TEM employed FEI Tecnai G2 F-30 TEM with a Schottky field-emission electron gun operated at 100 kV accelerating voltage. The TEM sample was prepared by collecting Si NCs directly onto a copper lacey carbon grid in the plasma reactor. 

FTIR measurements were performed using a Bruker Alpha IR spectrometer equipped with a diffuse reflectance (DRIFTS) accessory with a deuterated triglycine sulfate (DTGS) detector. All spectra were recorded from 375 to 7000 cm$^{-1}$ at 2 cm$^{-1}$ resolution, and averaged over 20 scans. 

The P incorporation in Si NCs was quantified by ICP-OES. Si NCs were digested in a mixture of hydrochloric acid (HCl), nitric acid (HNO$_{3}$) and hydrofluoric acid (HF). The elemental analysis was calibrated by the standards of Si and P samples.

Lateral two-terminal devices were fabricated on SiO$_{2}$ substrates with prepatterned Au interdigitated electrodes inside a nitrogen-filled glovebox. The spacing of electrodes is 30 $\mu$m, and the aspect ratio is 5317. The substrates were precleaned by sequential ultrasonication for 10 min each in acetone, methanol and isopropyl alcohol, and were treated in UV/Ozone for 20 min. As-produced Si NC powders were dissolved in anhydrous 1,2-dichlorobenzene (DCB), and cloudy stable suspensions were formed by ultrasonication. Si NC films were spin-coated from dispersions of 10 mg ml$^{-1}$.

Previous work has shown that low temperature annealing leads to an increase in the free electron concentration of the P-doped Si NCs and this is primarily attributed to the reduction of dangling bond defects during the annealing process \cite{rowe_phosphorus-doped_2013}. We notice that the annealed Si NC films exhibit higher conductance and improved stability compared with fresh-made films. For this study, all P-doped Si NC films were annealed at 125 \textdegree C overnight inside the glovebox before measurement. The O$_{2}$ level was controlled less than 0.1 ppm to minimize the oxidation of NCs during annealing. The devices were then transferred into another nitrogen-filled glovebox for subsequent electrical measurements. All handling and testing of devices was performed without air exposure.

The current-voltage ({\it I-V}) characteristics of the NC films were recorded in a Desert Cryogenics (Lakeshore) probe station in a nitrogen-filled glovebox with Keithley 236 and 237 source measuring units and homemade LabVIEW programs. Low temperature measurements employed a Lakeshore 331 temperature controller with a fixed ramp rate of 4 K/min. All electrical measurements were carried out in the dark and under vacuum at the pressure of $\sim$ 10$^{-3}$ Torr. 

%

\newpage
\clearpage

\section{Supplementary information:\\
Metal-insulator transition in films of doped semiconductor nanocrystals}

\renewcommand{\theequation}{S\arabic{equation}} 
\renewcommand{\thefigure}{S\arabic{figure}}%

\subsection{SI1 Geometry of $b$-contact \label{sec:b_contact}}

\begin{figure}[h!]
\includegraphics[width=.45\textwidth]{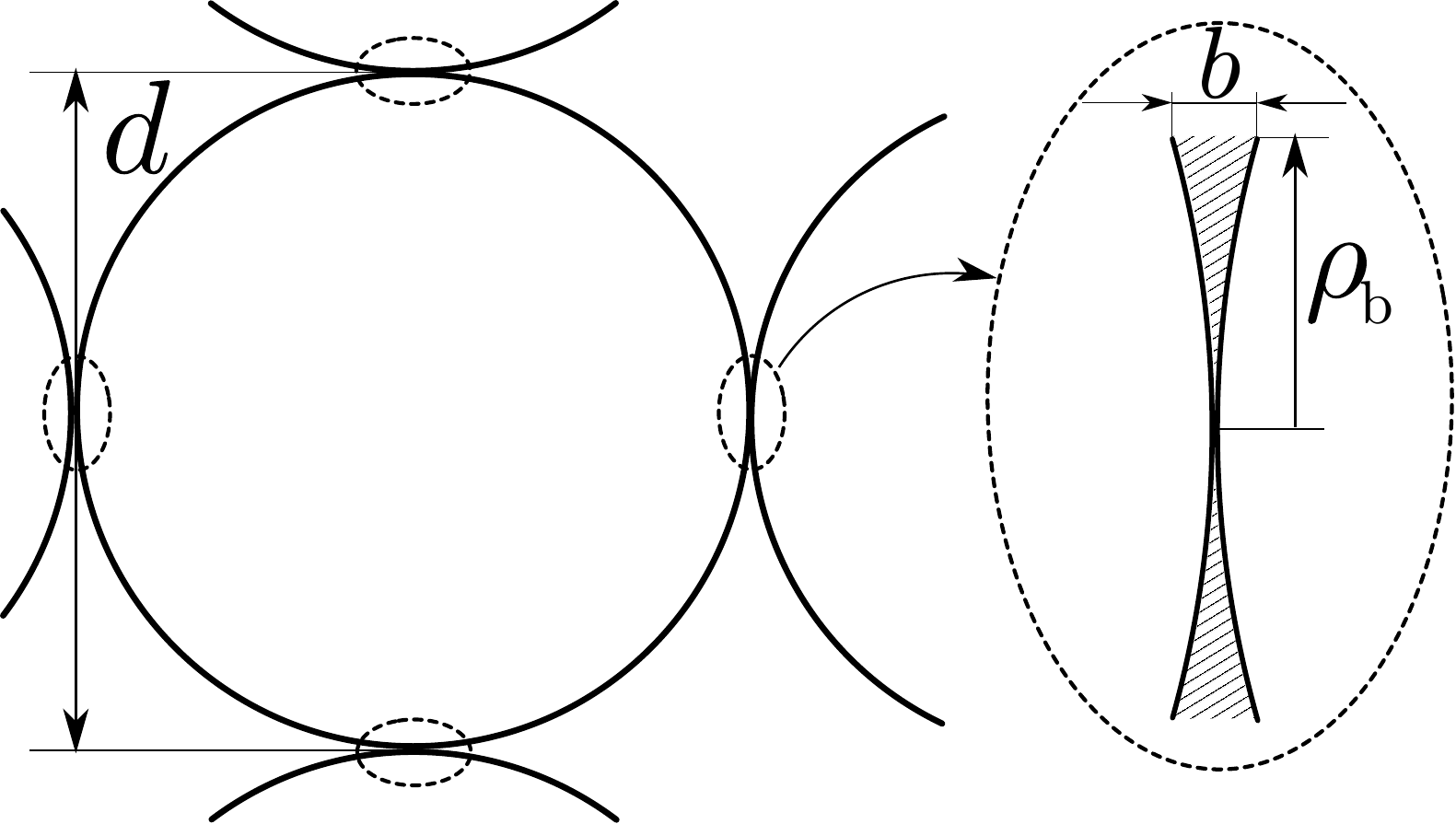}\\
\caption{Two NCs touching away from  facets are shown schematically. In this case, electrons tunnel through the $b$-contact which is depicted in the inset. The radius of the contact is $\rho_b=\sqrt{db/2}$, where $b$ is the decay length of an electron in the  medium surrounding NCs, which is shaded.}\label{fig:b_contact}
\end{figure}

In Fig. \ref{fig:b_contact} we schematically show the geometry of the $b$-contact, which determines the conductivity in the case when NCs touch each other away from facets.

\subsection{SI2 Proof of the universality of the MIT criterion Eq. (5)}
\label{sec:Anderson}
In the paper we used the criterion for the MIT Eq. (5):
\begin{equation}
  \label{eq:critical_G}
  G = \frac{e^2}{\pi \hbar}
\end{equation}

\begin{figure}[h]
\includegraphics[width=.5\textwidth]{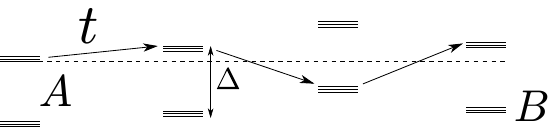}\\
\caption{Schematic illustration of the origin of the probability amplitude  to find an electron localized on the NC $A$  on the distant  NC $B$. The dashed line shows the energy of the tunneling electron. Each NC has $2l+1$  degenerate levels with the gap  $\Delta$ between them.}\label{fig:tunneling}
\end{figure}

 It specifies the  conductance between NCs at which the Anderson localization is eliminated. One can wonder why large energy gaps $\Delta$ between consecutive shells  are not making the Anderson delocalization more difficult in the degenerate case than in the non-degenerate one. Indeed, due to random charges of surrounding NCs the whole spectrum of a NC is shifted up and down with respect to the spectrum  of  nearest-neighbor NCs and the Fermi level of the system (see Fig. \ref{fig:tunneling}). The reason for the universality in this case is that a large number of degenerate levels in shells closest to the Fermi level in both contacting NCs compensates the large value of $\Delta$. Below we verify this compensation generalizing arguments of Ref. ~\cite{Anderson}   to the degenerate case.

The overlap integral between levels of two nearest neighbor NCs $$t = \int \Psi_1 \frac{\hbar^2}{2m^*} \Delta  \Psi_2^{*} dV,$$ where $\Psi_1,\Psi_2$ are wave functions on two contacting NCs. Following Ref. \cite{Anderson} we calculate  the amplitude to find an electron on the NC $B$ when it is localized on the NC $A$. The electron tunnels from $A$ to $B$  following a particular path with $S$ intermediate states in sequential NCs (see Fig. \ref{fig:tunneling} and  Ref. \cite{Anderson}).  The corresponding amplitude is
\begin{equation}
A\sim \left(\frac{t}{\Delta}\right)^{S+1},
\end{equation}
where $\Delta$ stands for the estimate of the absolute value of differences  between the energy of the tunneling electron and the (degenerate) energy levels of an intermediate NC. Due to the $2l+1$ degeneracy  of each  energy shell the total number of such paths is $(2l+1)^S$. Since the intermediate energy is the same, these tunneling amplitudes have the same sign. Adding all of them we get
\begin{equation}
A_{total}\sim \left(\frac{t}{\Delta}\right)^{S+1}(2l+1)^S\sim \left(\frac{t(2l+1)}{\Delta}\right)^S.
\end{equation}

This leads to the tunneling probability
\begin{equation}
P\sim \left(\frac{|t| (2l+1)}{\Delta}\right)^{2S}.
\end{equation}
When $S \rightarrow \infty$ the tunneling probability  exponentially diverges if $|t|(2l+1)/\Delta >1$. This means the delocalization of the electron. We can define a quantity $D=(2l+1)/\Delta$ as the density of states participating in the tunneling process within each dot. Then the universal  criterion for the MIT is 

\begin{equation}
  \label{eq:the_true_criterion}
|t|D \simeq 1.  
\end{equation}

The contact conductance between two quantum dots according to Ref. \cite{Glazman} is
\begin{equation}
G\simeq\frac{e^2}{h}|t|^2D^2.
\end{equation}
We find that even for the case of degenerate levels the   metal-insulator transition happens at the critical value of  $G$ which is determined by Eq. (\ref{eq:critical_G}).

We see that  the criterion for the MIT, Eq.~(\ref{eq:critical_G}), does not depend on the degeneracy of levels and is correct for both $a_B \ll d$ and $a_B \gg d $.

\subsection{SI3 Characterization of Si NCs}
\label{sec:characterization}

\begin{figure}[t!]
\centering
\includegraphics[width=0.5 \textwidth]{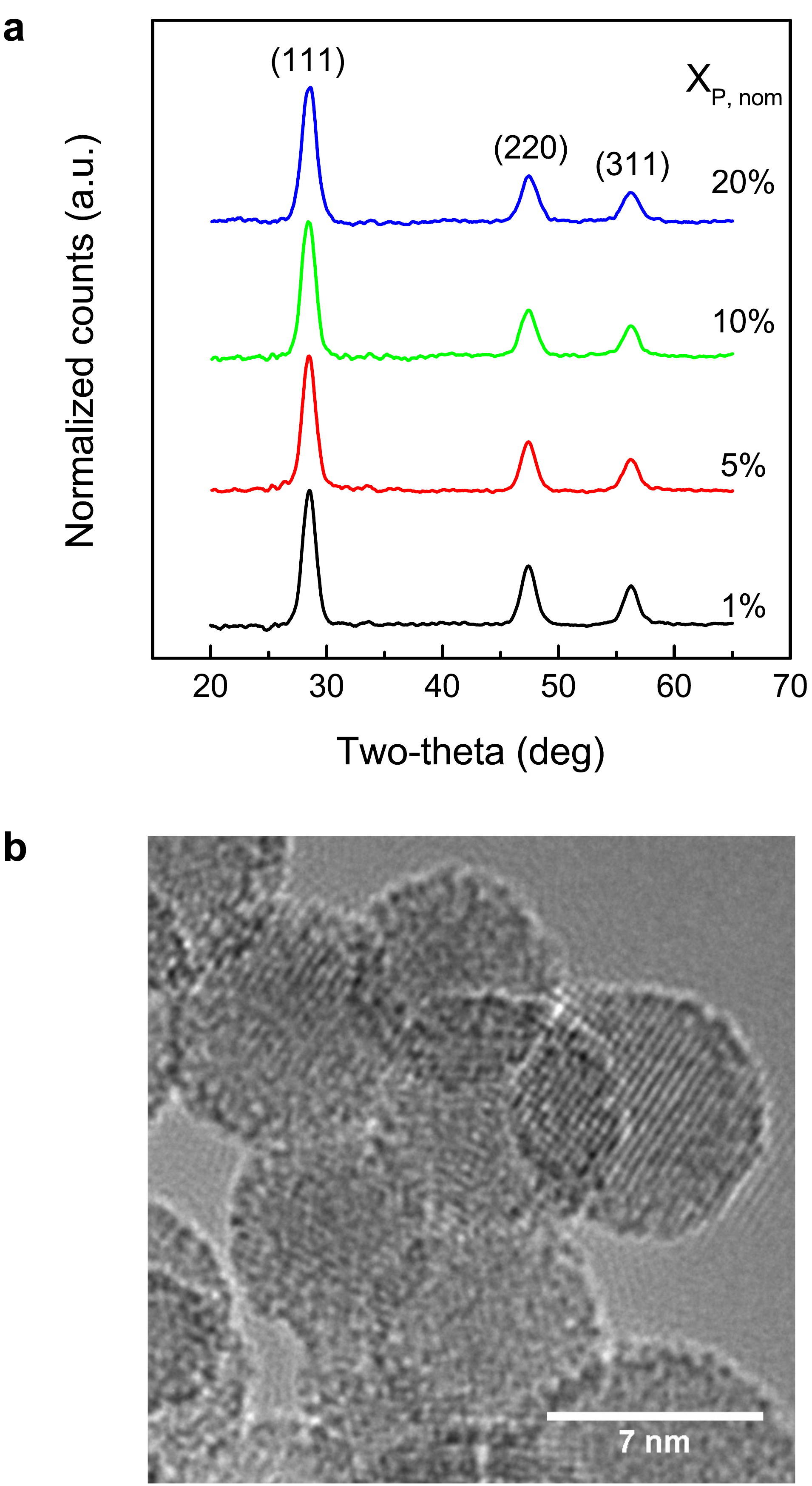} 

\caption{Structural characterization for P-doped Si NCs. {\bf a} XRD spectra for P-doped Si NCs at {\it X}$_{P,nom}$ = 1\%, 5\%, 10\% and 20\%. {\bf b} High resolution TEM image of nominal 10\% P-doped Si NCs.}
\label{fig:size}
\end{figure}

We used X-ray diffraction (XRD) to estimate the diameters of NCs. Figure \ref{fig:size}a shows well-defined XRD peaks corresponding to diamond cubic structure in P-doped Si NCs with nominal doping concentration {\it X}$_{P,nom}$ from 1\% to 20\%. The diameters of NCs shown in Table I are calculated from the peak broadening in XRD spectra with spherical correction \cite{borchert_determination_2005}. As {\it X}$_{P,nom}$ increases from 1\% to 20\%, the NC diameter decreases from 8.1 nm to 7.1 nm. This size reduction is likely caused by H$_{2}$ etching \cite{glass_jr._reaction_1996}, since PH$_{3}$ is diluted in H$_{2}$ with a volume fraction 15\%. 

A typical bright field transmission electron microscopy (TEM) image for nominal 10\% P-doped Si NCs is shown in Figure \ref{fig:size}b. The P doping does not alter the spherical shape of Si NCs and the NC diameter dispersion is $\sim$ 15\%  for all NCs used in this study. With the diameter of NCs and the concentration of free electrons $n$, we can calculate the average number of  electrons per NC $N$. For $X_{P,nom}=20$ \%, we get $N \sim 50$ electrons per NC.  All known parameters of Si NCs are summarized in Table I.

\subsection{SI4 Oxidation}
\label{sec:oxidation}
In previous work, we found that oxidation follows the Cabrera-Mott mechanism with a characteristic time $t_{m}=14.4$ min \cite{pereira_oxidation_2011}. Based on this oxidation mechanism, we can investigate the dependence of the localization length on the separation $s$ between NCs.

\begin{figure}[t!]
\centering
\includegraphics[width=0.5 \textwidth]{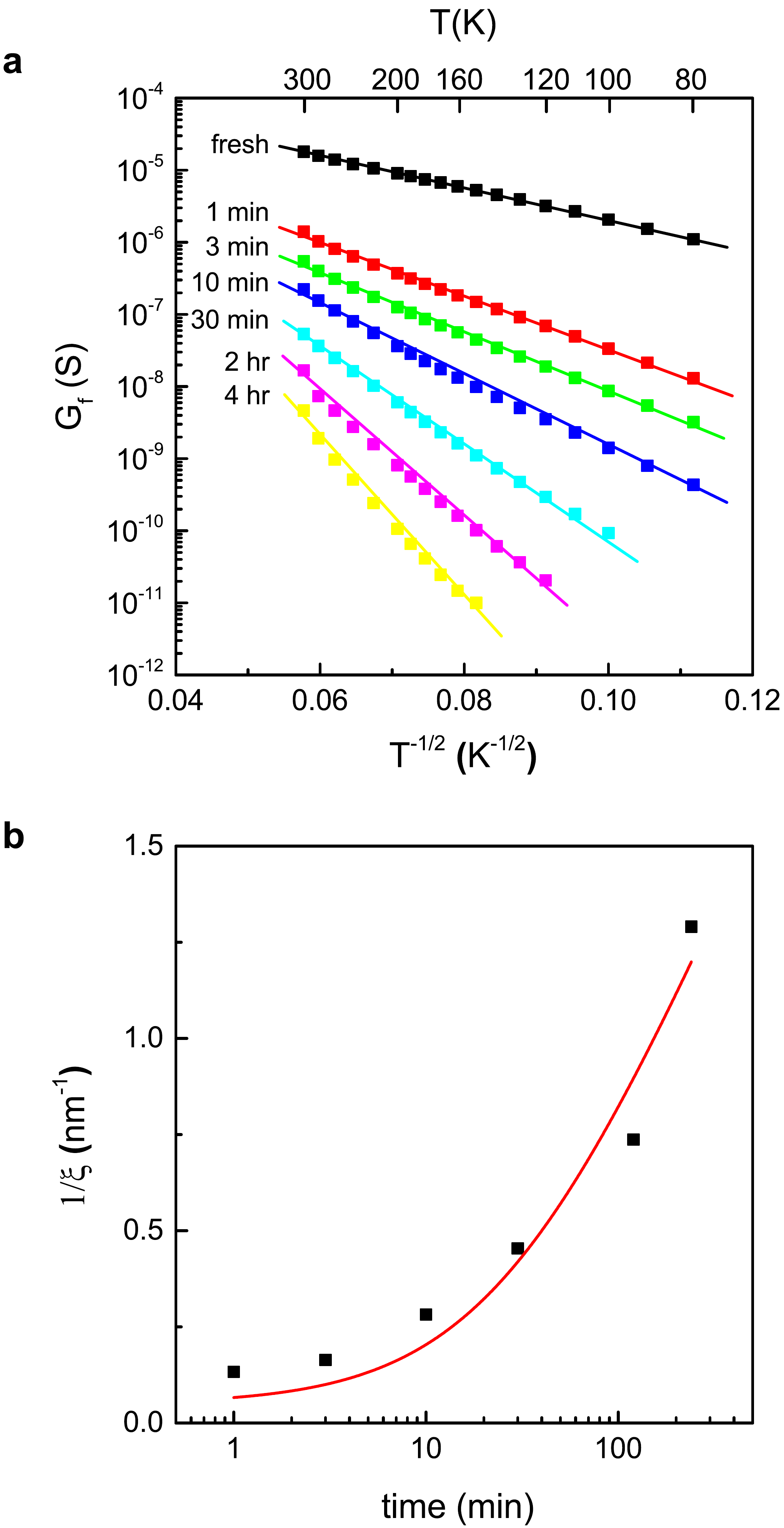} 

\caption{{\bf a} Temperature dependence of the ohmic conductance for films exposed to air from 1 min to 4 hrs. The solid lines are linear fits for each curve. {\bf b} Localization length $\xi$ of Si NC films vs oxidation time. Error bar for $\xi$ comes from the uncertainty caused by the linear fit and it is the same as the symbol size. The solid red line is the fit using Eq. \eqref{eq:Elovich}.}

\label{fig:oxidation}
\end{figure}

The 10\% P-doped Si NC film was exposed to the air for different periods of time. Temperature dependence of the ohmic conductance for films exposed to air from 1 min to 4 hrs is plotted against {\it T}$^{-1/2}$ in Figure \ref{fig:oxidation}a. The film conductance decreases with decreasing temperature for all measurements, and ES variable range hopping was observed for the majority of temperature range.  $T_{ES}$ is getting larger with increasing air exposure time, which indicates the decrease of the localization length with oxidation. 

This effect can be understood as follows \cite{skinner_theory_2012}. In ES variable range hopping, when an electron tunnels to a distant, non-neighboring NC at a distance {\it x}, its tunneling trajectory involves passing through a chain of intermediate NCs, and the decay of the electron wave function is dominated by passage through  gaps between neighboring NCs along the chain. As a consequence, the wave function is suppressed by a factor of $\sim \exp[-s x/b d]$ \cite{zhang_density_2004}, so that the localization length is given by $\xi \simeq  b  d / s$. The oxide growth on the Si NC surface increases $s$ and reduces $\xi$. 

The dynamics of the oxide growth on NC surface can be characterized by the Elovich equation \cite{aharoni_kinetics_1970}: $s =r_{0}t_{m}\ln\left ( 1 + t/t_{m} \right )$, where $r_{0}$ and $t_{m}$ are reaction rate and characteristic time, respectively. Now we have,

\begin{equation}
\label{eq:Elovich}
\frac{1}{\xi} = \frac{r_{0}t_{m}} {b d} \ln\left( 1+\frac{t}{t_{m}} \right ).
\end{equation}

The inverse localization length for oxidized Si NC films is plotted against air exposure time in a linear-log scale as shown in Figure \ref{fig:oxidation}b, and the red solid line is the fit with Eq. \eqref{eq:Elovich}. The experimental data are in good agreement with Cabrera-Mott oxidation mechanism. The characteristic time $t_m$ for P-doped Si NCs is found to be 28 min, twice larger than the intrinsic H-terminated Si NCs \cite{pereira_oxidation_2011}. This means faster oxidation of P-doped Si NCs.
In Cabrera-Mott mechanism, an electron from the cleaved Si-Si bond is transferred to an adsorbed O$_{2}$ molecule and the resulting O$_{2}^{-}$ ion drifts toward the cleaved Si-Si bond with assistance of the electrostatic potential \cite{cerofolini_model_2006}. Since oxidation of Si NCs requires electron tunneling,  faster oxidation is expected in $n$-type doped Si NCs \cite{pi_doping_2008}, as we see in our system.


\begin{thebibliography}{44}%
\makeatletter
\providecommand \@ifxundefined [1]{%
 \@ifx{#1\undefined}
}%
\providecommand \@ifnum [1]{%
 \ifnum #1\expandafter \@firstoftwo
 \else \expandafter \@secondoftwo
 \fi
}%
\providecommand \@ifx [1]{%
 \ifx #1\expandafter \@firstoftwo
 \else \expandafter \@secondoftwo
 \fi
}%
\providecommand \natexlab [1]{#1}%
\providecommand \enquote  [1]{``#1''}%
\providecommand \bibnamefont  [1]{#1}%
\providecommand \bibfnamefont [1]{#1}%
\providecommand \citenamefont [1]{#1}%
\providecommand \href@noop [0]{\@secondoftwo}%
\providecommand \href [0]{\begingroup \@sanitize@url \@href}%
\providecommand \@href[1]{\@@startlink{#1}\@@href}%
\providecommand \@@href[1]{\endgroup#1\@@endlink}%
\providecommand \@sanitize@url [0]{\catcode `\\12\catcode `\$12\catcode
  `\&12\catcode `\#12\catcode `\^12\catcode `\_12\catcode `\%12\relax}%
\providecommand \@@startlink[1]{}%
\providecommand \@@endlink[0]{}%
\providecommand \url  [0]{\begingroup\@sanitize@url \@url }%
\providecommand \@url [1]{\endgroup\@href {#1}{\urlprefix }}%
\providecommand \urlprefix  [0]{URL }%
\providecommand \Eprint [0]{\href }%
\providecommand \doibase [0]{http://dx.doi.org/}%
\providecommand \selectlanguage [0]{\@gobble}%
\providecommand \bibinfo  [0]{\@secondoftwo}%
\providecommand \bibfield  [0]{\@secondoftwo}%
\providecommand \translation [1]{[#1]}%
\providecommand \BibitemOpen [0]{}%
\providecommand \bibitemStop [0]{}%
\providecommand \bibitemNoStop [0]{.\EOS\space}%
\providecommand \EOS [0]{\spacefactor3000\relax}%
\providecommand \BibitemShut  [1]{\csname bibitem#1\endcsname}%
\let\auto@bib@innerbib\@empty
\bibitem [{\citenamefont {Gur}\ \emph {et~al.}(2005)\citenamefont {Gur},
  \citenamefont {Fromer}, \citenamefont {Geier},\ and\ \citenamefont
  {Alivisatos}}]{gur_air-stable_2005}%
  \BibitemOpen
  \bibfield  {author} {\bibinfo {author} {\bibfnamefont {I.}~\bibnamefont
  {Gur}}, \bibinfo {author} {\bibfnamefont {N.~A.}\ \bibnamefont {Fromer}},
  \bibinfo {author} {\bibfnamefont {M.~L.}\ \bibnamefont {Geier}}, \ and\
  \bibinfo {author} {\bibfnamefont {A.~P.}\ \bibnamefont {Alivisatos}},\ }\href
  {\doibase 10.1126/science.1117908} {\bibfield  {journal} {\bibinfo  {journal}
  {Science}\ }\textbf {\bibinfo {volume} {310}},\ \bibinfo {pages} {462}
  (\bibinfo {year} {2005})}\BibitemShut {NoStop}%
\bibitem [{\citenamefont {Wood}\ and\ \citenamefont
  {Bulovic}(2010)}]{wood_colloidal_2010}%
  \BibitemOpen
  \bibfield  {author} {\bibinfo {author} {\bibfnamefont {V.}~\bibnamefont
  {Wood}}\ and\ \bibinfo {author} {\bibfnamefont {V.}~\bibnamefont {Bulovic}},\
  }\href {\doibase 10.3402/nano.v1i0.5202} {\bibfield  {journal} {\bibinfo
  {journal} {Nano Reviews}\ }\textbf {\bibinfo {volume} {1}},\ \bibinfo {pages}
  {1} (\bibinfo {year} {2010})}\BibitemShut {NoStop}%
\bibitem [{\citenamefont {Turk}\ \emph {et~al.}(2014)\citenamefont {Turk},
  \citenamefont {Choi}, \citenamefont {Oh}, \citenamefont {Fafarman},
  \citenamefont {Diroll}, \citenamefont {Murray}, \citenamefont {Kagan},\ and\
  \citenamefont {Kikkawa}}]{turk_gate-induced_2014}%
  \BibitemOpen
  \bibfield  {author} {\bibinfo {author} {\bibfnamefont {M.~E.}\ \bibnamefont
  {Turk}}, \bibinfo {author} {\bibfnamefont {J.-H.}\ \bibnamefont {Choi}},
  \bibinfo {author} {\bibfnamefont {S.~J.}\ \bibnamefont {Oh}}, \bibinfo
  {author} {\bibfnamefont {A.~T.}\ \bibnamefont {Fafarman}}, \bibinfo {author}
  {\bibfnamefont {B.~T.}\ \bibnamefont {Diroll}}, \bibinfo {author}
  {\bibfnamefont {C.~B.}\ \bibnamefont {Murray}}, \bibinfo {author}
  {\bibfnamefont {C.~R.}\ \bibnamefont {Kagan}}, \ and\ \bibinfo {author}
  {\bibfnamefont {J.~M.}\ \bibnamefont {Kikkawa}},\ }\href {\doibase
  10.1021/nl5029655} {\bibfield  {journal} {\bibinfo  {journal} {Nano Letters}\
  }\textbf {\bibinfo {volume} {14}},\ \bibinfo {pages} {5948} (\bibinfo {year}
  {2014})}\BibitemShut {NoStop}%
\bibitem [{\citenamefont {Gresback}\ \emph {et~al.}(2014)\citenamefont
  {Gresback}, \citenamefont {Kramer}, \citenamefont {Ding}, \citenamefont
  {Chen}, \citenamefont {Kortshagen},\ and\ \citenamefont
  {Nozaki}}]{gresback_controlled_2014}%
  \BibitemOpen
  \bibfield  {author} {\bibinfo {author} {\bibfnamefont {R.}~\bibnamefont
  {Gresback}}, \bibinfo {author} {\bibfnamefont {N.~J.}\ \bibnamefont
  {Kramer}}, \bibinfo {author} {\bibfnamefont {Y.}~\bibnamefont {Ding}},
  \bibinfo {author} {\bibfnamefont {T.}~\bibnamefont {Chen}}, \bibinfo {author}
  {\bibfnamefont {U.~R.}\ \bibnamefont {Kortshagen}}, \ and\ \bibinfo {author}
  {\bibfnamefont {T.}~\bibnamefont {Nozaki}},\ }\href {\doibase
  10.1021/nn500182b} {\bibfield  {journal} {\bibinfo  {journal} {ACS Nano}\
  }\textbf {\bibinfo {volume} {8}},\ \bibinfo {pages} {5650} (\bibinfo {year}
  {2014})}\BibitemShut {NoStop}%
\bibitem [{\citenamefont {Alivisatos}(1996)}]{alivisatos_semiconductor_1996}%
  \BibitemOpen
  \bibfield  {author} {\bibinfo {author} {\bibfnamefont {A.~P.}\ \bibnamefont
  {Alivisatos}},\ }\href {\doibase 10.1126/science.271.5251.933} {\bibfield
  {journal} {\bibinfo  {journal} {Science}\ }\textbf {\bibinfo {volume}
  {271}},\ \bibinfo {pages} {933} (\bibinfo {year} {1996})}\BibitemShut
  {NoStop}%
\bibitem [{\citenamefont {Sargent}(2009)}]{sargent_infrared_2009}%
  \BibitemOpen
  \bibfield  {author} {\bibinfo {author} {\bibfnamefont {E.~H.}\ \bibnamefont
  {Sargent}},\ }\href {\doibase 10.1038/nphoton.2009.89} {\bibfield  {journal}
  {\bibinfo  {journal} {Nature Photonics}\ }\textbf {\bibinfo {volume} {3}},\
  \bibinfo {pages} {325} (\bibinfo {year} {2009})}\BibitemShut {NoStop}%
\bibitem [{\citenamefont {Murray}\ \emph {et~al.}(1993)\citenamefont {Murray},
  \citenamefont {Norris},\ and\ \citenamefont
  {Bawendi}}]{murray_synthesis_1993}%
  \BibitemOpen
  \bibfield  {author} {\bibinfo {author} {\bibfnamefont {C.~B.}\ \bibnamefont
  {Murray}}, \bibinfo {author} {\bibfnamefont {D.~J.}\ \bibnamefont {Norris}},
  \ and\ \bibinfo {author} {\bibfnamefont {M.~G.}\ \bibnamefont {Bawendi}},\
  }\href {\doibase 10.1021/ja00072a025} {\bibfield  {journal} {\bibinfo
  {journal} {Journal of the American Chemical Society}\ }\textbf {\bibinfo
  {volume} {115}},\ \bibinfo {pages} {8706} (\bibinfo {year}
  {1993})}\BibitemShut {NoStop}%
\bibitem [{\citenamefont {Talapin}\ and\ \citenamefont
  {Murray}(2005)}]{talapin_pbse_2005}%
  \BibitemOpen
  \bibfield  {author} {\bibinfo {author} {\bibfnamefont {D.~V.}\ \bibnamefont
  {Talapin}}\ and\ \bibinfo {author} {\bibfnamefont {C.~B.}\ \bibnamefont
  {Murray}},\ }\href {\doibase 10.1126/science.1116703} {\bibfield  {journal}
  {\bibinfo  {journal} {Science}\ }\textbf {\bibinfo {volume} {310}},\ \bibinfo
  {pages} {86} (\bibinfo {year} {2005})}\BibitemShut {NoStop}%
\bibitem [{\citenamefont {Yu}\ \emph {et~al.}(2003)\citenamefont {Yu},
  \citenamefont {Wang},\ and\ \citenamefont {Guyot-Sionnest}}]{yu_n-type_2003}%
  \BibitemOpen
  \bibfield  {author} {\bibinfo {author} {\bibfnamefont {D.}~\bibnamefont
  {Yu}}, \bibinfo {author} {\bibfnamefont {C.}~\bibnamefont {Wang}}, \ and\
  \bibinfo {author} {\bibfnamefont {P.}~\bibnamefont {Guyot-Sionnest}},\ }\href
  {\doibase 10.1126/science.1084424} {\bibfield  {journal} {\bibinfo  {journal}
  {Science}\ }\textbf {\bibinfo {volume} {300}},\ \bibinfo {pages} {1277}
  (\bibinfo {year} {2003})}\BibitemShut {NoStop}%
\bibitem [{\citenamefont {Wang}\ \emph {et~al.}(2001)\citenamefont {Wang},
  \citenamefont {Shim},\ and\ \citenamefont
  {Guyot-Sionnest}}]{wang_electrochromic_2001}%
  \BibitemOpen
  \bibfield  {author} {\bibinfo {author} {\bibfnamefont {C.}~\bibnamefont
  {Wang}}, \bibinfo {author} {\bibfnamefont {M.}~\bibnamefont {Shim}}, \ and\
  \bibinfo {author} {\bibfnamefont {P.}~\bibnamefont {Guyot-Sionnest}},\ }\href
  {\doibase 10.1126/science.291.5512.2390} {\bibfield  {journal} {\bibinfo
  {journal} {Science}\ }\textbf {\bibinfo {volume} {291}},\ \bibinfo {pages}
  {2390} (\bibinfo {year} {2001})}\BibitemShut {NoStop}%
\bibitem [{\citenamefont {Oh}\ \emph {et~al.}(2013)\citenamefont {Oh},
  \citenamefont {Berry}, \citenamefont {Choi}, \citenamefont {Gaulding},
  \citenamefont {Paik}, \citenamefont {Hong}, \citenamefont {Murray},\ and\
  \citenamefont {Kagan}}]{oh_stoichiometric_2013}%
  \BibitemOpen
  \bibfield  {author} {\bibinfo {author} {\bibfnamefont {S.~J.}\ \bibnamefont
  {Oh}}, \bibinfo {author} {\bibfnamefont {N.~E.}\ \bibnamefont {Berry}},
  \bibinfo {author} {\bibfnamefont {J.-H.}\ \bibnamefont {Choi}}, \bibinfo
  {author} {\bibfnamefont {E.~A.}\ \bibnamefont {Gaulding}}, \bibinfo {author}
  {\bibfnamefont {T.}~\bibnamefont {Paik}}, \bibinfo {author} {\bibfnamefont
  {S.-H.}\ \bibnamefont {Hong}}, \bibinfo {author} {\bibfnamefont {C.~B.}\
  \bibnamefont {Murray}}, \ and\ \bibinfo {author} {\bibfnamefont {C.~R.}\
  \bibnamefont {Kagan}},\ }\href {\doibase 10.1021/nn3057356} {\bibfield
  {journal} {\bibinfo  {journal} {ACS Nano}\ }\textbf {\bibinfo {volume} {7}},\
  \bibinfo {pages} {2413} (\bibinfo {year} {2013})}\BibitemShut {NoStop}%
\bibitem [{\citenamefont {Norris}\ \emph {et~al.}(2008)\citenamefont {Norris},
  \citenamefont {Efros},\ and\ \citenamefont {Erwin}}]{norris_doped_2008}%
  \BibitemOpen
  \bibfield  {author} {\bibinfo {author} {\bibfnamefont {D.~J.}\ \bibnamefont
  {Norris}}, \bibinfo {author} {\bibfnamefont {A.~L.}\ \bibnamefont {Efros}}, \
  and\ \bibinfo {author} {\bibfnamefont {S.~C.}\ \bibnamefont {Erwin}},\ }\href
  {\doibase 10.1126/science.1143802} {\bibfield  {journal} {\bibinfo  {journal}
  {Science}\ }\textbf {\bibinfo {volume} {319}},\ \bibinfo {pages} {1776}
  (\bibinfo {year} {2008})}\BibitemShut {NoStop}%
\bibitem [{\citenamefont {Mocatta}\ \emph {et~al.}(2011)\citenamefont
  {Mocatta}, \citenamefont {Cohen}, \citenamefont {Schattner}, \citenamefont
  {Millo}, \citenamefont {Rabani},\ and\ \citenamefont
  {Banin}}]{mocatta_heavily_2011}%
  \BibitemOpen
  \bibfield  {author} {\bibinfo {author} {\bibfnamefont {D.}~\bibnamefont
  {Mocatta}}, \bibinfo {author} {\bibfnamefont {G.}~\bibnamefont {Cohen}},
  \bibinfo {author} {\bibfnamefont {J.}~\bibnamefont {Schattner}}, \bibinfo
  {author} {\bibfnamefont {O.}~\bibnamefont {Millo}}, \bibinfo {author}
  {\bibfnamefont {E.}~\bibnamefont {Rabani}}, \ and\ \bibinfo {author}
  {\bibfnamefont {U.}~\bibnamefont {Banin}},\ }\href {\doibase
  10.1126/science.1196321} {\bibfield  {journal} {\bibinfo  {journal}
  {Science}\ }\textbf {\bibinfo {volume} {332}},\ \bibinfo {pages} {77}
  (\bibinfo {year} {2011})}\BibitemShut {NoStop}%
\bibitem [{\citenamefont {Sahu}\ \emph {et~al.}(2012)\citenamefont {Sahu},
  \citenamefont {Kang}, \citenamefont {Kompch}, \citenamefont {Notthoff},
  \citenamefont {Wills}, \citenamefont {Deng}, \citenamefont {Winterer},
  \citenamefont {Frisbie},\ and\ \citenamefont
  {Norris}}]{sahu_electronic_2012}%
  \BibitemOpen
  \bibfield  {author} {\bibinfo {author} {\bibfnamefont {A.}~\bibnamefont
  {Sahu}}, \bibinfo {author} {\bibfnamefont {M.~S.}\ \bibnamefont {Kang}},
  \bibinfo {author} {\bibfnamefont {A.}~\bibnamefont {Kompch}}, \bibinfo
  {author} {\bibfnamefont {C.}~\bibnamefont {Notthoff}}, \bibinfo {author}
  {\bibfnamefont {A.~W.}\ \bibnamefont {Wills}}, \bibinfo {author}
  {\bibfnamefont {D.}~\bibnamefont {Deng}}, \bibinfo {author} {\bibfnamefont
  {M.}~\bibnamefont {Winterer}}, \bibinfo {author} {\bibfnamefont {C.~D.}\
  \bibnamefont {Frisbie}}, \ and\ \bibinfo {author} {\bibfnamefont {D.~J.}\
  \bibnamefont {Norris}},\ }\href {\doibase 10.1021/nl300880g} {\bibfield
  {journal} {\bibinfo  {journal} {Nano Letters}\ }\textbf {\bibinfo {volume}
  {12}},\ \bibinfo {pages} {2587} (\bibinfo {year} {2012})}\BibitemShut
  {NoStop}%
\bibitem [{\citenamefont
  {Guyot-Sionnest}(2012)}]{guyot-sionnest_electrical_2012}%
  \BibitemOpen
  \bibfield  {author} {\bibinfo {author} {\bibfnamefont {P.}~\bibnamefont
  {Guyot-Sionnest}},\ }\href {\doibase 10.1021/jz300048y} {\bibfield  {journal}
  {\bibinfo  {journal} {The Journal of Physical Chemistry Letters}\ }\textbf
  {\bibinfo {volume} {3}},\ \bibinfo {pages} {1169} (\bibinfo {year}
  {2012})}\BibitemShut {NoStop}%
\bibitem [{\citenamefont {Shabaev}\ \emph {et~al.}(2013)\citenamefont
  {Shabaev}, \citenamefont {Efros},\ and\ \citenamefont
  {Efros}}]{shabaev_dark_2013}%
  \BibitemOpen
  \bibfield  {author} {\bibinfo {author} {\bibfnamefont {A.}~\bibnamefont
  {Shabaev}}, \bibinfo {author} {\bibfnamefont {A.~L.}\ \bibnamefont {Efros}},
  \ and\ \bibinfo {author} {\bibfnamefont {A.~L.}\ \bibnamefont {Efros}},\
  }\href {\doibase 10.1021/nl403033f} {\bibfield  {journal} {\bibinfo
  {journal} {Nano Letters}\ }\textbf {\bibinfo {volume} {13}},\ \bibinfo
  {pages} {5454} (\bibinfo {year} {2013})}\BibitemShut {NoStop}%
\bibitem [{\citenamefont {Scheele}(2015)}]{Band_like_transport_review}%
  \BibitemOpen
  \bibfield  {author} {\bibinfo {author} {\bibfnamefont {M.}~\bibnamefont
  {Scheele}},\ }\href {\doibase 10.1515/zpch-2014-0587} {\bibfield  {journal}
  {\bibinfo  {journal} {Zeitschrift f\"{u}r Physikalische Chemie}\ }\textbf
  {\bibinfo {volume} {229}},\ \bibinfo {pages} {167} (\bibinfo {year}
  {2015})}\BibitemShut {NoStop}%
\bibitem [{\citenamefont {Mott}(1968)}]{mott_metal-insulator_1968}%
  \BibitemOpen
  \bibfield  {author} {\bibinfo {author} {\bibfnamefont {N.~F.}\ \bibnamefont
  {Mott}},\ }\href {\doibase 10.1103/RevModPhys.40.677} {\bibfield  {journal}
  {\bibinfo  {journal} {Reviews of Modern Physics}\ }\textbf {\bibinfo {volume}
  {40}},\ \bibinfo {pages} {677} (\bibinfo {year} {1968})}\BibitemShut
  {NoStop}%
\bibitem [{\citenamefont {Fu}\ \emph {et~al.}(2016)\citenamefont {Fu},
  \citenamefont {Reich},\ and\ \citenamefont
  {Shklovskii}}]{localization_length_NC}%
  \BibitemOpen
  \bibfield  {author} {\bibinfo {author} {\bibfnamefont {H.}~\bibnamefont
  {Fu}}, \bibinfo {author} {\bibfnamefont {K.~V.}\ \bibnamefont {Reich}}, \
  and\ \bibinfo {author} {\bibfnamefont {B.~I.}\ \bibnamefont {Shklovskii}},\
  }\href {\doibase 10.1103/PhysRevB.93.125430} {\bibfield  {journal} {\bibinfo
  {journal} {Phys. Rev. B}\ }\textbf {\bibinfo {volume} {93}},\ \bibinfo
  {pages} {125430} (\bibinfo {year} {2016})}\BibitemShut {NoStop}%
\bibitem [{\citenamefont {Sharvin}(1965)}]{Sharvin}%
  \BibitemOpen
  \bibfield  {author} {\bibinfo {author} {\bibfnamefont {Y.~V.}\ \bibnamefont
  {Sharvin}},\ }\href@noop {} {\bibfield  {journal} {\bibinfo  {journal}
  {Soviet Journal of Experimental and Theoretical Physics}\ }\textbf {\bibinfo
  {volume} {21}},\ \bibinfo {pages} {655} (\bibinfo {year} {1965})}\BibitemShut
  {NoStop}%
\bibitem [{\citenamefont {Nikoli\ifmmode~\acute{c}\else \'{c}\fi{}}\ and\
  \citenamefont {Allen}(1999)}]{transport_through_constriction}%
  \BibitemOpen
  \bibfield  {author} {\bibinfo {author} {\bibfnamefont {B.}~\bibnamefont
  {Nikoli\ifmmode~\acute{c}\else \'{c}\fi{}}}\ and\ \bibinfo {author}
  {\bibfnamefont {P.~B.}\ \bibnamefont {Allen}},\ }\href {\doibase
  10.1103/PhysRevB.60.3963} {\bibfield  {journal} {\bibinfo  {journal} {Phys.
  Rev. B}\ }\textbf {\bibinfo {volume} {60}},\ \bibinfo {pages} {3963}
  (\bibinfo {year} {1999})}\BibitemShut {NoStop}%
\bibitem [{\citenamefont {Nazarov}\ and\ \citenamefont
  {Blanter}(2009)}]{nazarov2009quantum}%
  \BibitemOpen
  \bibfield  {author} {\bibinfo {author} {\bibfnamefont {Y.}~\bibnamefont
  {Nazarov}}\ and\ \bibinfo {author} {\bibfnamefont {Y.}~\bibnamefont
  {Blanter}},\ }\href@noop {} {\emph {\bibinfo {title} {Quantum Transport:
  Introduction to Nanoscience}}}\ (\bibinfo  {publisher} {Cambridge University
  Press},\ \bibinfo {year} {2009})\BibitemShut {NoStop}%
\bibitem [{\citenamefont {Matveev}(1995)}]{Matveev}%
  \BibitemOpen
  \bibfield  {author} {\bibinfo {author} {\bibfnamefont {K.~A.}\ \bibnamefont
  {Matveev}},\ }\href {\doibase 10.1103/PhysRevB.51.1743} {\bibfield  {journal}
  {\bibinfo  {journal} {Phys. Rev. B}\ }\textbf {\bibinfo {volume} {51}},\
  \bibinfo {pages} {1743} (\bibinfo {year} {1995})}\BibitemShut {NoStop}%
\bibitem [{\citenamefont {Beloborodov}\ \emph {et~al.}(2007)\citenamefont
  {Beloborodov}, \citenamefont {Lopatin}, \citenamefont {Vinokur},\ and\
  \citenamefont {Efetov}}]{beloborodov_granular_2007}%
  \BibitemOpen
  \bibfield  {author} {\bibinfo {author} {\bibfnamefont {I.~S.}\ \bibnamefont
  {Beloborodov}}, \bibinfo {author} {\bibfnamefont {A.~V.}\ \bibnamefont
  {Lopatin}}, \bibinfo {author} {\bibfnamefont {V.~M.}\ \bibnamefont
  {Vinokur}}, \ and\ \bibinfo {author} {\bibfnamefont {K.~B.}\ \bibnamefont
  {Efetov}},\ }\href {\doibase 10.1103/RevModPhys.79.469} {\bibfield  {journal}
  {\bibinfo  {journal} {Reviews of Modern Physics}\ }\textbf {\bibinfo {volume}
  {79}},\ \bibinfo {pages} {469} (\bibinfo {year} {2007})}\BibitemShut
  {NoStop}%
\bibitem [{\citenamefont {Liljeroth}\ \emph {et~al.}(2006)\citenamefont
  {Liljeroth}, \citenamefont {Overgaag}, \citenamefont {Urbieta}, \citenamefont
  {Grandidier}, \citenamefont {Hickey},\ and\ \citenamefont
  {Vanmaekelbergh}}]{facets}%
  \BibitemOpen
  \bibfield  {author} {\bibinfo {author} {\bibfnamefont {P.}~\bibnamefont
  {Liljeroth}}, \bibinfo {author} {\bibfnamefont {K.}~\bibnamefont {Overgaag}},
  \bibinfo {author} {\bibfnamefont {A.}~\bibnamefont {Urbieta}}, \bibinfo
  {author} {\bibfnamefont {B.}~\bibnamefont {Grandidier}}, \bibinfo {author}
  {\bibfnamefont {S.~G.}\ \bibnamefont {Hickey}}, \ and\ \bibinfo {author}
  {\bibfnamefont {D.}~\bibnamefont {Vanmaekelbergh}},\ }\href {\doibase
  10.1103/PhysRevLett.97.096803} {\bibfield  {journal} {\bibinfo  {journal}
  {Phys. Rev. Lett.}\ }\textbf {\bibinfo {volume} {97}},\ \bibinfo {pages}
  {096803} (\bibinfo {year} {2006})}\BibitemShut {NoStop}%
\bibitem [{\citenamefont {Lee}\ \emph {et~al.}(2011)\citenamefont {Lee},
  \citenamefont {Kovalenko}, \citenamefont {Huang}, \citenamefont {Chung},\
  and\ \citenamefont {Talapin}}]{lee_band-like_2011}%
  \BibitemOpen
  \bibfield  {author} {\bibinfo {author} {\bibfnamefont {J.-S.}\ \bibnamefont
  {Lee}}, \bibinfo {author} {\bibfnamefont {M.~V.}\ \bibnamefont {Kovalenko}},
  \bibinfo {author} {\bibfnamefont {J.}~\bibnamefont {Huang}}, \bibinfo
  {author} {\bibfnamefont {D.~S.}\ \bibnamefont {Chung}}, \ and\ \bibinfo
  {author} {\bibfnamefont {D.~V.}\ \bibnamefont {Talapin}},\ }\href {\doibase
  10.1038/nnano.2011.46} {\bibfield  {journal} {\bibinfo  {journal} {Nature
  Nanotechnology}\ }\textbf {\bibinfo {volume} {6}},\ \bibinfo {pages} {348}
  (\bibinfo {year} {2011})}\BibitemShut {NoStop}%
\bibitem [{\citenamefont {Choi}\ \emph {et~al.}(2012)\citenamefont {Choi},
  \citenamefont {Fafarman}, \citenamefont {Oh}, \citenamefont {Ko},
  \citenamefont {Kim}, \citenamefont {Diroll}, \citenamefont {Muramoto},
  \citenamefont {Gillen}, \citenamefont {Murray},\ and\ \citenamefont
  {Kagan}}]{choi_bandlike_2012}%
  \BibitemOpen
  \bibfield  {author} {\bibinfo {author} {\bibfnamefont {J.-H.}\ \bibnamefont
  {Choi}}, \bibinfo {author} {\bibfnamefont {A.~T.}\ \bibnamefont {Fafarman}},
  \bibinfo {author} {\bibfnamefont {S.~J.}\ \bibnamefont {Oh}}, \bibinfo
  {author} {\bibfnamefont {D.-K.}\ \bibnamefont {Ko}}, \bibinfo {author}
  {\bibfnamefont {D.~K.}\ \bibnamefont {Kim}}, \bibinfo {author} {\bibfnamefont
  {B.~T.}\ \bibnamefont {Diroll}}, \bibinfo {author} {\bibfnamefont
  {S.}~\bibnamefont {Muramoto}}, \bibinfo {author} {\bibfnamefont {J.~G.}\
  \bibnamefont {Gillen}}, \bibinfo {author} {\bibfnamefont {C.~B.}\
  \bibnamefont {Murray}}, \ and\ \bibinfo {author} {\bibfnamefont {C.~R.}\
  \bibnamefont {Kagan}},\ }\href {\doibase 10.1021/nl301104z} {\bibfield
  {journal} {\bibinfo  {journal} {Nano Letters}\ }\textbf {\bibinfo {volume}
  {12}},\ \bibinfo {pages} {2631} (\bibinfo {year} {2012})}\BibitemShut
  {NoStop}%
\bibitem [{\citenamefont {Liu}\ \emph {et~al.}(2013)\citenamefont {Liu},
  \citenamefont {Lee},\ and\ \citenamefont
  {Talapin}}]{Talapin_2013_high_mobility}%
  \BibitemOpen
  \bibfield  {author} {\bibinfo {author} {\bibfnamefont {W.}~\bibnamefont
  {Liu}}, \bibinfo {author} {\bibfnamefont {J.-S.}\ \bibnamefont {Lee}}, \ and\
  \bibinfo {author} {\bibfnamefont {D.~V.}\ \bibnamefont {Talapin}},\ }\href
  {\doibase 10.1021/ja308200f} {\bibfield  {journal} {\bibinfo  {journal}
  {Journal of the American Chemical Society}\ }\textbf {\bibinfo {volume}
  {135}},\ \bibinfo {pages} {1349} (\bibinfo {year} {2013})}\BibitemShut
  {NoStop}%
\bibitem [{\citenamefont {Skinner}\ \emph {et~al.}(2012)\citenamefont
  {Skinner}, \citenamefont {Chen},\ and\ \citenamefont
  {Shklovskii}}]{skinner_theory_2012}%
  \BibitemOpen
  \bibfield  {author} {\bibinfo {author} {\bibfnamefont {B.}~\bibnamefont
  {Skinner}}, \bibinfo {author} {\bibfnamefont {T.}~\bibnamefont {Chen}}, \
  and\ \bibinfo {author} {\bibfnamefont {B.~I.}\ \bibnamefont {Shklovskii}},\
  }\href {\doibase 10.1103/PhysRevB.85.205316} {\bibfield  {journal} {\bibinfo
  {journal} {Physical Review B}\ }\textbf {\bibinfo {volume} {85}},\ \bibinfo
  {pages} {205316} (\bibinfo {year} {2012})}\BibitemShut {NoStop}%
\bibitem [{\citenamefont {Mangolini}\ \emph {et~al.}(2005)\citenamefont
  {Mangolini}, \citenamefont {Thimsen},\ and\ \citenamefont
  {Kortshagen}}]{mangolini_high-yield_2005}%
  \BibitemOpen
  \bibfield  {author} {\bibinfo {author} {\bibfnamefont {L.}~\bibnamefont
  {Mangolini}}, \bibinfo {author} {\bibfnamefont {E.}~\bibnamefont {Thimsen}},
  \ and\ \bibinfo {author} {\bibfnamefont {U.}~\bibnamefont {Kortshagen}},\
  }\href {\doibase 10.1021/nl050066y} {\bibfield  {journal} {\bibinfo
  {journal} {Nano Letters}\ }\textbf {\bibinfo {volume} {5}},\ \bibinfo {pages}
  {655} (\bibinfo {year} {2005})}\BibitemShut {NoStop}%
\bibitem [{\citenamefont {Marra}\ \emph {et~al.}(1998)\citenamefont {Marra},
  \citenamefont {Edelberg}, \citenamefont {Naone},\ and\ \citenamefont
  {Aydil}}]{modes_phonon_Si}%
  \BibitemOpen
  \bibfield  {author} {\bibinfo {author} {\bibfnamefont {D.~C.}\ \bibnamefont
  {Marra}}, \bibinfo {author} {\bibfnamefont {E.~A.}\ \bibnamefont {Edelberg}},
  \bibinfo {author} {\bibfnamefont {R.~L.}\ \bibnamefont {Naone}}, \ and\
  \bibinfo {author} {\bibfnamefont {E.~S.}\ \bibnamefont {Aydil}},\ }\href
  {\doibase 10.1116/1.581520} {\bibfield  {journal} {\bibinfo  {journal}
  {Journal of Vacuum Science and Technology A}\ }\textbf {\bibinfo {volume}
  {16}},\ \bibinfo {pages} {3199} (\bibinfo {year} {1998})}\BibitemShut
  {NoStop}%
\bibitem [{\citenamefont {Miura}\ \emph {et~al.}(1996)\citenamefont {Miura},
  \citenamefont {Niwano}, \citenamefont {Shoji},\ and\ \citenamefont
  {Miyamoto}}]{Si_H_modes}%
  \BibitemOpen
  \bibfield  {author} {\bibinfo {author} {\bibfnamefont {T.}~\bibnamefont
  {Miura}}, \bibinfo {author} {\bibfnamefont {M.}~\bibnamefont {Niwano}},
  \bibinfo {author} {\bibfnamefont {D.}~\bibnamefont {Shoji}}, \ and\ \bibinfo
  {author} {\bibfnamefont {N.}~\bibnamefont {Miyamoto}},\ }\href {\doibase
  10.1063/1.362670} {\bibfield  {journal} {\bibinfo  {journal} {Journal of
  Applied Physics}\ }\textbf {\bibinfo {volume} {79}},\ \bibinfo {pages} {4373}
  (\bibinfo {year} {1996})}\BibitemShut {NoStop}%
\bibitem [{\citenamefont {Rowe}\ \emph {et~al.}(2013)\citenamefont {Rowe},
  \citenamefont {Jeong}, \citenamefont {Mkhoyan},\ and\ \citenamefont
  {Kortshagen}}]{rowe_phosphorus-doped_2013}%
  \BibitemOpen
  \bibfield  {author} {\bibinfo {author} {\bibfnamefont {D.~J.}\ \bibnamefont
  {Rowe}}, \bibinfo {author} {\bibfnamefont {J.~S.}\ \bibnamefont {Jeong}},
  \bibinfo {author} {\bibfnamefont {K.~A.}\ \bibnamefont {Mkhoyan}}, \ and\
  \bibinfo {author} {\bibfnamefont {U.~R.}\ \bibnamefont {Kortshagen}},\ }\href
  {\doibase 10.1021/nl4001184} {\bibfield  {journal} {\bibinfo  {journal} {Nano
  Letters}\ }\textbf {\bibinfo {volume} {13}},\ \bibinfo {pages} {1317}
  (\bibinfo {year} {2013})}\BibitemShut {NoStop}%
\bibitem [{\citenamefont {Zabrodskii}(2001)}]{zabrodskii_coulomb_2001}%
  \BibitemOpen
  \bibfield  {author} {\bibinfo {author} {\bibfnamefont {A.~G.}\ \bibnamefont
  {Zabrodskii}},\ }\href {\doibase 10.1080/13642810108205796} {\bibfield
  {journal} {\bibinfo  {journal} {Philosophical Magazine Part B}\ }\textbf
  {\bibinfo {volume} {81}},\ \bibinfo {pages} {1131} (\bibinfo {year}
  {2001})}\BibitemShut {NoStop}%
\bibitem [{\citenamefont {Maxwell}(1881)}]{maxwell_treatise_1881}%
  \BibitemOpen
  \bibfield  {author} {\bibinfo {author} {\bibfnamefont {J.~C.}\ \bibnamefont
  {Maxwell}},\ }\href
  {http://books.google.com/books?hl=en&lr=&id=WtExAQAAMAAJ&oi=fnd&pg=PR1&dq=J.+C.+Maxwell,+A+Treatise+on+Electricity+and+Magnetism&ots=jD5ZJrOwzk&sig=zxQaC0n6Fqoit7lxTWX49dmVvkE}
  {\emph {\bibinfo {title} {A treatise on electricity and magnetism}}},\
  Vol.~\bibinfo {volume} {1}\ (\bibinfo  {publisher} {Clarendon press},\
  \bibinfo {year} {1881})\BibitemShut {NoStop}%
\bibitem [{\citenamefont {Anderson}(1972)}]{Anderson}%
  \BibitemOpen
  \bibfield  {author} {\bibinfo {author} {\bibfnamefont {P.}~\bibnamefont
  {Anderson}},\ }\href@noop {} {\bibfield  {journal} {\bibinfo  {journal}
  {Proceedings of the National Academy of Sciences of the United States of
  America}\ }\textbf {\bibinfo {volume} {69}},\ \bibinfo {pages} {1097}
  (\bibinfo {year} {1972})}\BibitemShut {NoStop}%
\bibitem [{\citenamefont {Glazman}\ and\ \citenamefont
  {Pustilnik}(2005)}]{Glazman}%
  \BibitemOpen
  \bibfield  {author} {\bibinfo {author} {\bibfnamefont {L.~I.}\ \bibnamefont
  {Glazman}}\ and\ \bibinfo {author} {\bibfnamefont {M.}~\bibnamefont
  {Pustilnik}},\ }\href {\doibase 10.1016/S0924-8099(05)80050-2} {\emph
  {\bibinfo {title} {Nanophysics: Coherence and Transport}}},\ edited by\
  \bibinfo {editor} {\bibfnamefont {H.}~\bibnamefont {Bouchiat}}, \bibinfo
  {editor} {\bibfnamefont {Y.}~\bibnamefont {Gefen}}, \bibinfo {editor}
  {\bibfnamefont {S.}~\bibnamefont {Gueron}}, \bibinfo {editor} {\bibfnamefont
  {G.}~\bibnamefont {Montambaux}}, \ and\ \bibinfo {editor} {\bibfnamefont
  {J.}~\bibnamefont {Dalibard}},\ \bibinfo {series} {Les Houches},
  Vol.~\bibinfo {volume} {81}\ (\bibinfo  {publisher} {Elsevier},\ \bibinfo
  {year} {2005})\ pp.\ \bibinfo {pages} {427 -- 478}\BibitemShut {NoStop}%
\bibitem [{\citenamefont {Borchert}\ \emph {et~al.}(2005)\citenamefont
  {Borchert}, \citenamefont {Shevchenko}, \citenamefont {Robert}, \citenamefont
  {Mekis}, \citenamefont {Kornowski}, \citenamefont {Grubel},\ and\
  \citenamefont {Weller}}]{borchert_determination_2005}%
  \BibitemOpen
  \bibfield  {author} {\bibinfo {author} {\bibfnamefont {H.}~\bibnamefont
  {Borchert}}, \bibinfo {author} {\bibfnamefont {E.~V.}\ \bibnamefont
  {Shevchenko}}, \bibinfo {author} {\bibfnamefont {A.}~\bibnamefont {Robert}},
  \bibinfo {author} {\bibfnamefont {I.}~\bibnamefont {Mekis}}, \bibinfo
  {author} {\bibfnamefont {A.}~\bibnamefont {Kornowski}}, \bibinfo {author}
  {\bibfnamefont {G.}~\bibnamefont {Grubel}}, \ and\ \bibinfo {author}
  {\bibfnamefont {H.}~\bibnamefont {Weller}},\ }\href
  {http://pubs.acs.org/doi/abs/10.1021/la0477183} {\bibfield  {journal}
  {\bibinfo  {journal} {Langmuir}\ }\textbf {\bibinfo {volume} {21}},\ \bibinfo
  {pages} {1931} (\bibinfo {year} {2005})}\BibitemShut {NoStop}%
\bibitem [{\citenamefont {Glass~Jr.}\ \emph {et~al.}(1996)\citenamefont
  {Glass~Jr.}, \citenamefont {Wovchko},\ and\ \citenamefont
  {Yates~Jr.}}]{glass_jr._reaction_1996}%
  \BibitemOpen
  \bibfield  {author} {\bibinfo {author} {\bibfnamefont {J.~A.}\ \bibnamefont
  {Glass~Jr.}}, \bibinfo {author} {\bibfnamefont {E.~A.}\ \bibnamefont
  {Wovchko}}, \ and\ \bibinfo {author} {\bibfnamefont {J.~T.}\ \bibnamefont
  {Yates~Jr.}},\ }\href {\doibase 10.1016/0039-6028(95)01014-9} {\bibfield
  {journal} {\bibinfo  {journal} {Surface Science}\ }\textbf {\bibinfo {volume}
  {348}},\ \bibinfo {pages} {325} (\bibinfo {year} {1996})}\BibitemShut
  {NoStop}%
\bibitem [{\citenamefont {Pereira}\ \emph {et~al.}(2011)\citenamefont
  {Pereira}, \citenamefont {Rowe}, \citenamefont {Anthony},\ and\ \citenamefont
  {Kortshagen}}]{pereira_oxidation_2011}%
  \BibitemOpen
  \bibfield  {author} {\bibinfo {author} {\bibfnamefont {R.~N.}\ \bibnamefont
  {Pereira}}, \bibinfo {author} {\bibfnamefont {D.~J.}\ \bibnamefont {Rowe}},
  \bibinfo {author} {\bibfnamefont {R.~J.}\ \bibnamefont {Anthony}}, \ and\
  \bibinfo {author} {\bibfnamefont {U.}~\bibnamefont {Kortshagen}},\ }\href
  {\doibase 10.1103/PhysRevB.83.155327} {\bibfield  {journal} {\bibinfo
  {journal} {Physical Review B}\ }\textbf {\bibinfo {volume} {83}},\ \bibinfo
  {pages} {155327} (\bibinfo {year} {2011})}\BibitemShut {NoStop}%
\bibitem [{\citenamefont {Zhang}\ and\ \citenamefont
  {Shklovskii}(2004)}]{zhang_density_2004}%
  \BibitemOpen
  \bibfield  {author} {\bibinfo {author} {\bibfnamefont {J.}~\bibnamefont
  {Zhang}}\ and\ \bibinfo {author} {\bibfnamefont {B.~I.}\ \bibnamefont
  {Shklovskii}},\ }\href {\doibase 10.1103/PhysRevB.70.115317} {\bibfield
  {journal} {\bibinfo  {journal} {Physical Review B}\ }\textbf {\bibinfo
  {volume} {70}},\ \bibinfo {pages} {115317} (\bibinfo {year}
  {2004})}\BibitemShut {NoStop}%
\bibitem [{\citenamefont {Aharoni}\ and\ \citenamefont
  {Tompkins}(1970)}]{aharoni_kinetics_1970}%
  \BibitemOpen
  \bibfield  {author} {\bibinfo {author} {\bibfnamefont {C.}~\bibnamefont
  {Aharoni}}\ and\ \bibinfo {author} {\bibfnamefont {F.~C.}\ \bibnamefont
  {Tompkins}},\ }in\ \href
  {http://www.sciencedirect.com/science/article/pii/S0360056408605635} {\emph
  {\bibinfo {booktitle} {Advances in {Catalysis}}}},\ Vol.~\bibinfo {volume}
  {21},\ \bibinfo {editor} {edited by\ \bibinfo {editor} {\bibfnamefont {H.~P.
  a. P. B.~W.}\ \bibnamefont {D.D.~Eley}}}\ (\bibinfo  {publisher} {Academic
  Press},\ \bibinfo {year} {1970})\ pp.\ \bibinfo {pages} {1--49}\BibitemShut
  {NoStop}%
\bibitem [{\citenamefont {Cerofolini}\ \emph {et~al.}(2006)\citenamefont
  {Cerofolini}, \citenamefont {Mascolo},\ and\ \citenamefont
  {Vlad}}]{cerofolini_model_2006}%
  \BibitemOpen
  \bibfield  {author} {\bibinfo {author} {\bibfnamefont {G.~F.}\ \bibnamefont
  {Cerofolini}}, \bibinfo {author} {\bibfnamefont {D.}~\bibnamefont {Mascolo}},
  \ and\ \bibinfo {author} {\bibfnamefont {M.~O.}\ \bibnamefont {Vlad}},\
  }\href {\doibase 10.1063/1.2245191} {\bibfield  {journal} {\bibinfo
  {journal} {Journal of Applied Physics}\ }\textbf {\bibinfo {volume} {100}},\
  \bibinfo {pages} {054308} (\bibinfo {year} {2006})}\BibitemShut {NoStop}%
\bibitem [{\citenamefont {Pi}\ \emph {et~al.}(2008)\citenamefont {Pi},
  \citenamefont {Gresback}, \citenamefont {Liptak}, \citenamefont {Campbell},\
  and\ \citenamefont {Kortshagen}}]{pi_doping_2008}%
  \BibitemOpen
  \bibfield  {author} {\bibinfo {author} {\bibfnamefont {X.}~\bibnamefont
  {Pi}}, \bibinfo {author} {\bibfnamefont {R.}~\bibnamefont {Gresback}},
  \bibinfo {author} {\bibfnamefont {R.~W.}\ \bibnamefont {Liptak}}, \bibinfo
  {author} {\bibfnamefont {S.}~\bibnamefont {Campbell}}, \ and\ \bibinfo
  {author} {\bibfnamefont {U.}~\bibnamefont {Kortshagen}},\ }\href {\doibase
  10.1063/1.2897291} {\bibfield  {journal} {\bibinfo  {journal} {Applied
  Physics Letters}\ }\textbf {\bibinfo {volume} {92}},\ \bibinfo {pages}
  {123102} (\bibinfo {year} {2008})}\BibitemShut {NoStop}%
\end{thebibliography}

\end{document}